\documentclass[pss,fleqn,embeddedheads]{w-art}
\usepackage{times}
\usepackage{w-thm}
\usepackage{amssymb,amsmath,psfrag,url}

\newcommand{\MyField}[1]{{\bf{#1}}}

\newcommand{\MyTensor}[1]{{{{#1}}}}
\newcommand{\MySField}[1]{{{#1}}}
\newcommand{\MyMatrix}[1]{\mathrm{#1}}

\usepackage[]{graphicx}
\begin{document}
\DOIsuffix{theDOIsuffix}
\Volume{XX}
\Issue{1}
\Copyrightissue{01}
\Month{01}
\Year{2004}
\pagespan{1}{}
\subjclass[pacs]{07.05.Tp, 78.20.Bh, 42.70.Qs, 42.81.Qb}



\title[FEM simulation of radiation losses]{Finite Element simulation of radiation losses in photonic crystal fibers}


\author[J. Pomplun]{Jan Pomplun\footnote{Corresponding
     author: e-mail: {\sf pomplun@zib.de}, Phone: +49\,30\,841\,85\,273}}
\address{Zuse Institute Berlin,
Takustra{\ss}e 7,
14\,195 Berlin,
Germany}
\author[L. Zschiedrich]{Lin Zschiedrich}
\author[R. Klose]{Roland Klose}
\author[F. Schmidt]{Frank Schmidt}
\author[S. Burger]{Sven Burger}

\begin{abstract}
In our work we focus on the accurate computation of light propagation in finite size photonic crystal structures with the finite element method (FEM). We discuss how we utilize numerical concepts like high-order finite elements, transparent boundary conditions and goal-oriented error estimators for adaptive grid refinement in order to compute radiation leakage in photonic crystal fibers and waveguides. Due to the fast convergence of our method we can use it e.g. to optimize the design of photonic crystal structures with respect to geometrical parameters, to minimize radiation losses and to compute attenutation spectra for different geometries.
\end{abstract}
\maketitle                   

\noindent
Published in phys. stat. sol. (a) \textbf{204}, No. 11, 3822-3837 (2007). \\\noindent
URL: http://www.interscience.wiley.com/

\section{Introduction}
Photonic crystals consist of a periodic arrangement of materials with different refractive indices, usually a dielectric material and air with a period on the scale of the wavelength \cite{JOA95}. This structure allows to control light propagation on very short distances. Due to their small size photonic crystal structures can be used to fabricate integrated optical components like resonators and waveguides. Photonic crystal fibers \cite{CRE99,RUS03} are an important application example which utilizes the light guiding principles of photonic crystals. They are used e.g. in nonlinear applications \cite{BEN02} for high power transmission of light in optical astronomy. 

Radiation losses from PhC fibers are one of the loss mechanism which potentially restrict their technological application \cite{COU06}. They have therefore to be controlled by proper component design. Due to the complex geometry and large number of glass air interfaces the accurate computation of light propagation in photonic crystal structures is a challenging task \cite{POM07}. To achieve a realistic model of a real device its finite size and the exterior domain have to be taken into account. The coupling of a photonic crystal fiber to the exterior will always lead to unwanted losses which with a proper fiber design are very small. For numerical computation transparent boundary conditions are used to take into account the surrounding material. Even for simple geometries the accurate computation of radiation losses can become difficult and many numerical methods can fail especially for systems where these losses are very small \cite{Bienstman:06a}.

In this paper we review an approach to this problem using adaptive finite element algorithms and we apply this approach to accurate computation of radiation losses in hexgonal and kagome-structured PhC fibers. Computational results agree well with experimental transmission spectra measured in \cite{COU06}.
In the context of topics related to the DFG priority programme photonic crystals, we have applied adaptive finite element algorithms to a variety of nano-optical problems \cite{Burger2005a,Burger2006b,Burger2005w,Enkrich2005a,HOL06,Kalkbrenner2005a,LindenEDKZKSBSW06,POM07,ZSC07}.
\section{Formulation of propagation mode problem}
For the analysis of photonic crystal fibers we start with the derivation of the mathematical formulation of the propagation mode problem of the electric and magnetic field. The geometry of a waveguide system like a photonic crystal fiber is invariant in one spatial dimension along the fiber. Here we choose the $z$-direction. Then a propagating mode is a solution to the time harmonic Maxwell's equations with frequency $\omega$, which exhibits a harmonic dependency in $z$-direction:
\begin{eqnarray}
\MyField{E} & = & \MyField{E}_{\mathrm{pm}}(x, y)\exp \left(ik_{z}z\right)\nonumber\\
\MyField{H} & = & \MyField{H}_{\mathrm{pm}}(x, y)\exp \left(ik_{z}z\right)\label{eq:propAnsatz}. 
\end{eqnarray}
$\MyField{E}_{\mathrm{pm}}(x, y)$ and $\MyField{H}_{\mathrm{pm}}(x, y)$ are the electric and magnetic propagation modes and the parameter $k_{z}$ is called propagation constant. If the permittivity $\MyTensor{\epsilon}$ and permeability $\MyTensor{\mu}$ can be written as:
\begin{equation}
\MyTensor{\epsilon} =
\left[
\begin{array}{cc}
\MyTensor{\epsilon}_{\perp\,\perp} & 0\\
0 & \MyTensor{\epsilon}_{zz} 
\end{array}
\right]  
\quad \mbox{and} \quad 
\MyTensor{\mu} =
\left[
\begin{array}{cc}
\MyTensor{\mu}_{\perp\,\perp} & 0 \\
0 & \MyTensor{\mu}_{zz} \label{eq:permitPermea}
\end{array}
\right],
\end{equation}
we can split the propagation mode into a transversal and a longitudinal component:
\begin{equation}
\MyField{E}_{\mathrm{pm}}(x, y) = 
\left[
\begin{array}{c}\MyField{E}_{\perp}(x, y) \\ \MySField{E}_{z}(x, y) \end{array}
\right].\label{eq:hprop}
\end{equation}
Inserting (\ref{eq:propAnsatz}) with (\ref{eq:permitPermea}) and (\ref{eq:hprop}) into Maxwell's equations yields:
\begin{equation}
\left[
\begin{array}{cc}
\MyMatrix{P} \nabla_{\perp} \MyTensor{\mu}_{zz}^{-1} \nabla_{\perp} \cdot \MyMatrix{P} 
-k_z^2 \MyMatrix{P} \MyTensor{\mu}_{\perp\,\perp}^{-1} \MyMatrix{P}\, &
-ik_z \MyMatrix{P} \MyTensor{\mu}_{\perp\,\perp}^{-1} \MyMatrix{P} \nabla_{\perp} \\
-ik_z\nabla_{\perp}\cdot \MyMatrix{P} \MyTensor{\mu}_{\perp\,\perp}^{-1} \MyMatrix{P} &
\nabla_{\perp}\cdot \MyMatrix{P} \MyTensor{\mu}_{\perp\,\perp}^{-1} \MyMatrix{P} \nabla_{\perp}
\end{array}
\right]
\left[
\begin{array}{c}\MyField{E}_{\perp} \\ \MySField{E}_{z} \end{array}
\right] =
\left[
\begin{array}{cc}
\omega^{2}\MyTensor{\epsilon}_{\perp\,\perp} & 0 \\
0 & \omega^{2}\MyTensor{\epsilon}_{zz}
\end{array}
\right]
\left[
\begin{array}{c}\MyField{E}_{\perp} \\ \MySField{E}_{z} \end{array}
\right],\label{eq:evp}
\end{equation}
with
\begin{equation}
\MyMatrix{P} =
\left[
\begin{array}{cc}
0 & -1 \\
1 & 0 
\end{array}
\right]
, \quad 
\nabla_{\perp} = 
\left[
\begin{array}{c} 
\partial_x \\
\partial_y
\end{array} 
\right]. 
\end{equation}
Now we define $\tilde{\MySField{E}}_{z}=k_{z}\MySField{E}_{z}$ and get:
\begin{equation}
\MyMatrix{ A}
\left[
\begin{array}{c}
\MyField{E}_{\perp} \\ 
\tilde{\MySField{E}}_{z} 
\end{array}
\right]
=
k_z^2 \,
\MyMatrix{ B}
\left[
\begin{array}{c}
\MyField{E}_{\perp} \\ 
\tilde{\MySField{E}}_{z} 
\end{array}
\right]
\quad x \in {\mathbb R}^{2}.  \label{eq:evp1}
\end{equation}
with
\begin{eqnarray}
  \label{eq:ABMatrices}
  \MyMatrix{A}&=&\left[
\begin{array}{cc}
\MyMatrix{P} \nabla_{\perp} \MyTensor{\mu}_{zz}^{-1} \nabla_{\perp} \cdot \MyMatrix{P} 
-\omega^{2}\MyTensor{\epsilon}_{\perp\,\perp}\, &
-i \MyMatrix{P} \MyTensor{\mu}_{\perp\,\perp}^{-1} \MyMatrix{P} \nabla_{\perp} \\
0 &
\nabla_{\perp}\cdot \MyMatrix{P} \MyTensor{\mu}_{\perp\,\perp}^{-1} \MyMatrix{P} \nabla_{\perp}- \omega^{2}\MyTensor{\epsilon}_{zz}
\end{array}
\right]\\
\MyMatrix{B}&=&\left[
\begin{array}{cc}
\MyMatrix{P} \MyTensor{\mu}_{\perp\,\perp}^{-1} \MyMatrix{P}
 & 0 \\
i\nabla_{\perp}\cdot \MyMatrix{P} \MyTensor{\mu}_{\perp\,\perp}^{-1} \MyMatrix{P} &
0
\end{array}
\right]
\end{eqnarray}
Eq. (\ref{eq:evp1}) is a generalized eigenvalue problem for the propagation constant $k_{z}$ and propagation mode $\MyField{E}_{\mathrm{pm}}(x, y)$. We get a similar equation for the magnetic field $\MyField{H}_{\mathrm{pm}}(x, y)$ exchanging $\MyTensor{\epsilon}$ and $\MyTensor{\mu}$. For our numerical analysis we define the effective refractive index $n_{\mathrm{eff}}$ which we will also refer to as eigenvalue:
\begin{equation}
\label{eq:neff}
n_{\mathrm{eff}} = \frac{k_z}{k_0} \qquad \mbox{with} \quad k_0 = \frac{2\pi}{\lambda_0},
\end{equation}
where $\lambda_{0}$ is the vacuum wavelength of light.


\section{Discretization of Maxwell's equations with the Finite Element Method}
For the numerical solution of the propagation mode problem Eq. (\ref{eq:evp1}) derived in the previous section we use the finite element method \cite{MON03} which we sketch briefly in the following. We start with the curl curl equation for the electric field. Since we want to solve an eigenvalue problem we are looking for pairs $\MyField{E}$ and $k_{z}$ such that:
\begin{eqnarray}
  \label{eq:curlcurlE}
&&\nabla_{k_z}\times\frac{1}{\MyTensor{\mu}}\nabla_{k_z}\times\MyField E-\frac{\omega^{2}\MyTensor{\epsilon}}{c^{2}} \MyField{E} = 0\quad \mbox{in $\Omega$}  \\
&&\left(\frac{1}{\MyTensor{\mu}}\nabla_{k_z}\times\MyField{E}\right)\times\vec n=\MyField{F}
\quad\mbox{given on $\Gamma$ (Neumann boundary
condition)}\nonumber\\\label{boundary}
\end{eqnarray}
holds, with $\nabla_{k_{z}}=[\partial_{x},\partial_{y},ik_{z}]^{T}$. For application of the finite element method we have to derive a weak formulation of this equation. Therefore we multiply (\ref{eq:curlcurlE}) with a vectorial test function
$\MyField{\Phi} \in V=H(curl)$ \cite{MON03} and integrate over the domain $\Omega$:
\begin{eqnarray}
  \label{eq:MWweak1}
  \int_{\Omega}\left\{\overline{\MyField{\Phi}}\cdot\left[\nabla_{k_z}\times\frac{1}{\MyTensor{\mu}}\nabla_{k_z}\times\MyField{ E}\right]-\frac{\omega^{2}\MyTensor{\epsilon}}{c^{2}} \overline{\MyField{\Phi}}\cdot\MyField{ E}\right\}d^{3}r=0\,,\;\forall {\MyField{\Phi}}\in V,
\end{eqnarray}
where bar denotes complex conjugation. After a partial integration we arrive at the weak formulation of Maxwell's equations:\\
Find $\MyField{ E}\in V=H(curl)$ such that
\begin{eqnarray}
  \label{eq:MWweak2}
&&\int_{\Omega}\left\{\overline{\left(\nabla_{k_z}\times\MyField{\Phi}\right)}\cdot\left(\frac{1}{\MyTensor{\mu}}\nabla_{k_z}\times\MyField{ E}\right)-\frac{\omega^{2}\MyTensor{\epsilon}}{c^{2}} \overline{\MyField{\Phi}}\cdot\MyField{ E}\right\}d^{3}r=\int_{\Gamma}\overline{\MyField{\Phi}}\cdot\MyField{ F} d^{2}r\,,\;\forall \MyField{\Phi}\in V.
\end{eqnarray}
We define the following bilinear functionals:
\begin{eqnarray}
  \label{eq:forms}
a(\MyField w,\MyField v)&=&\int_{\Omega}\overline{\left(\MyField\nabla_{k_z}\times\MyField w\right)}\cdot\left(\frac{1}{\MyTensor{\mu}}\MyField\nabla_{k_z}\times\MyField v\right)-\frac{\MyTensor{\omega^{2}\epsilon}}{c^{2}}\overline{\MyField w}\cdot\MyField v \,d^{3}r,\label{defa}\\
f(\MyField w)&=&\int_{\Gamma}\overline{\MyField w}\cdot\MyField F d^{2}r\label{defF}
\end{eqnarray}
\begin{figure}[h]
(a)\hspace{5cm}(b)\hspace{5cm}(c)\\
\includegraphics[height=4.1cm]{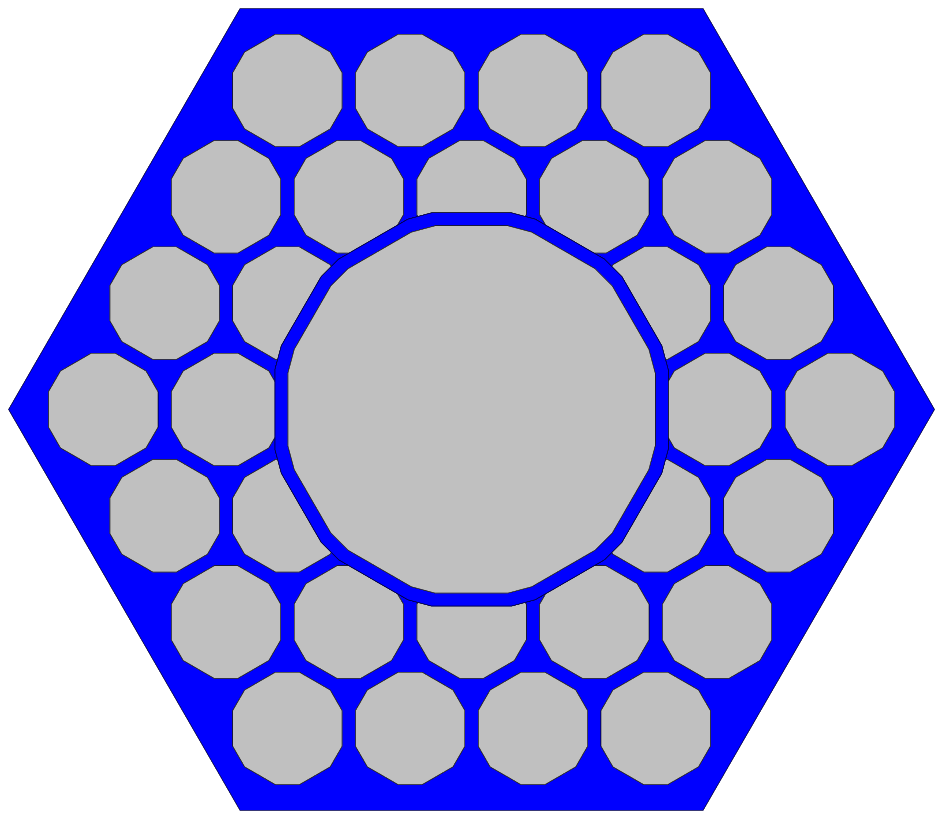}\hspace{0.3cm}
\includegraphics[height=4.1cm]{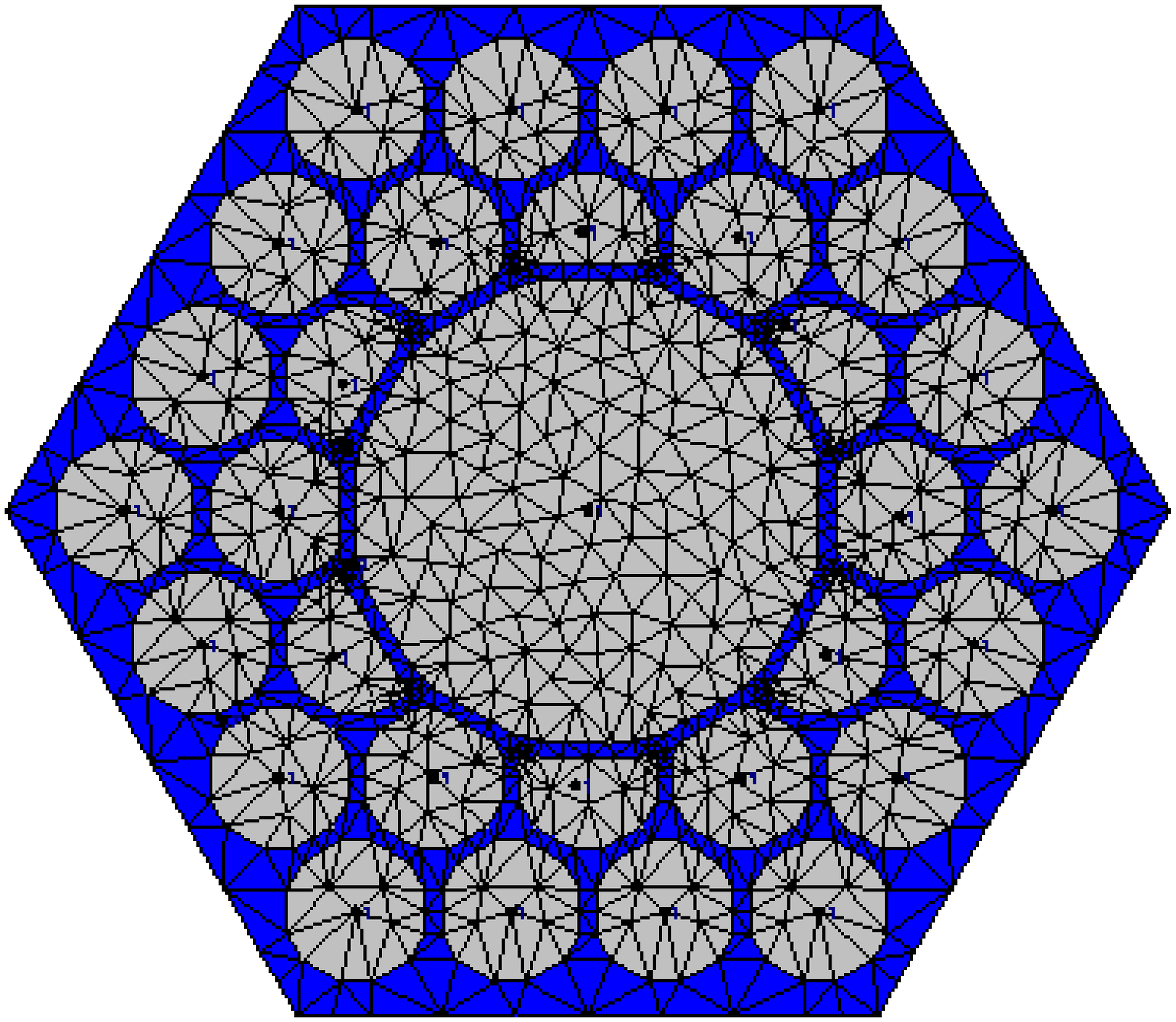}\hfill
\includegraphics[height=4.4cm]{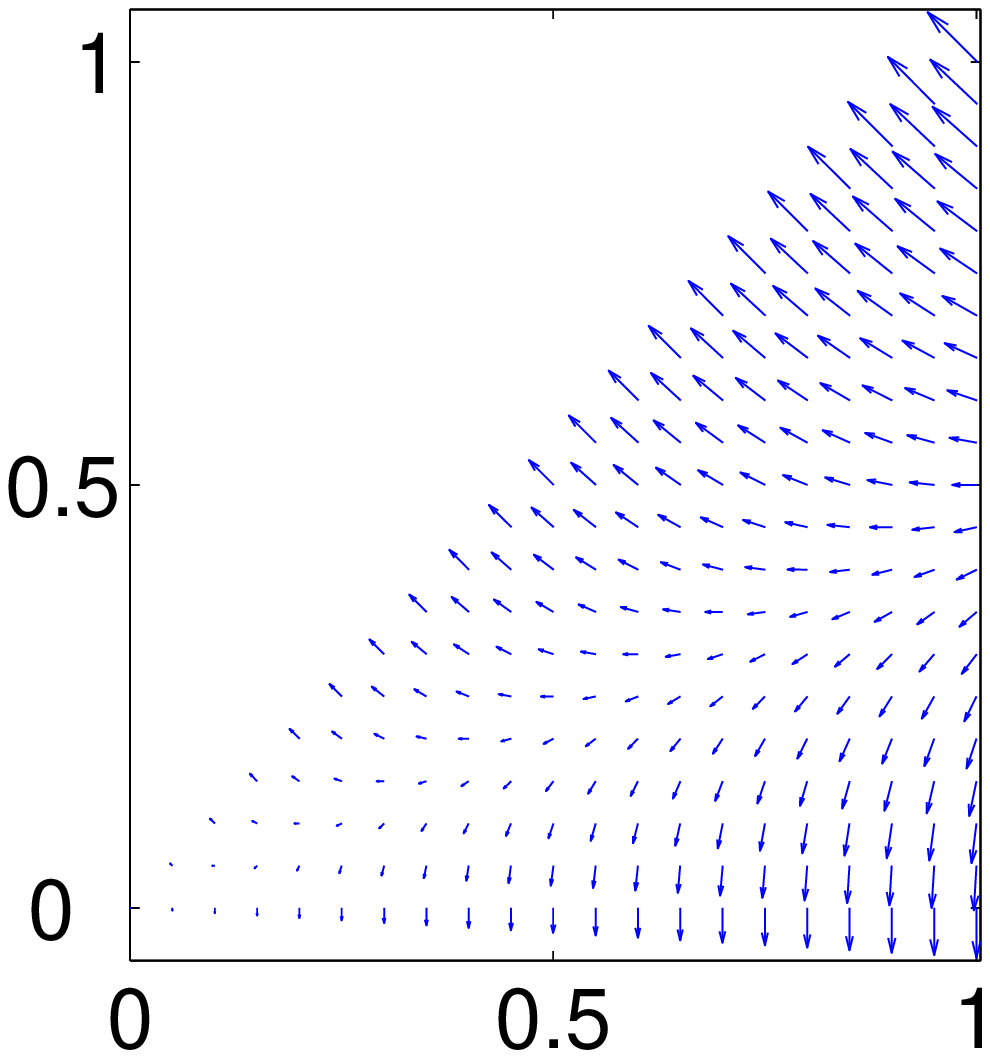}\hfill
\caption{\label{fig:fem}(a) Computational domain and (b) triangulation of hollow-core photonic crystal fiber; (c) example of vectorial ansatz function on a reference patch.}
\end{figure}
Now we discretize this equation by restricting the space $V$ to a finite dimensional subspace $V_{h}\subset V$, $dim V_{h}=N$. This subspace and therewith the approximate solution are constructed as follows. One starts with a computational domain $\Omega$, see Fig. \ref{fig:fem}(a). This domain is subdivided into small patches, e.g. triangles in 2D and tetrahedrons in 3D, Fig \ref{fig:fem}(b). On these patches vectorial ansatz functions $\MyField{\nu}_{{i}}$ are defined. Usually on each patch the ansatz funtions $\MyField{\nu}_{{i}}$ form a basis of a polynomial function space of a certain degree $p$ \cite{MON03}. The approximate solution ${\MyField{E}_{h}}$ for the electric field is a superposition of these ansatz functions of all patches:
\begin{eqnarray}
  \label{eq:EAnsatz}
  \MyField{E}_{h}=\sum_{i=1}^{N}a_{i}\MyField{\nu}_{i}
\end{eqnarray}
Together with (\ref{defa}), (\ref{defF}) the discrete version of Maxwell's equations (\ref{eq:MWweak2}) reads:
\begin{eqnarray}
  \label{eq:MWweak3}
&&  \sum_{i=1}^{N}a_{i} a(\nu_{i},\nu_{j})=f(\nu_{j})\,,\;\forall j=1,\dots,N
\end{eqnarray}
which is a linear system of equations for the unknown coefficients $a_{i}$:
\begin{eqnarray}
  \label{eq:MWweak4}
&&\MyMatrix{A}\cdot \vec a=\vec f\;,\nonumber\\
\mbox{with}&&\MyMatrix A_{ij}=a(\nu_{i},\nu_{j}),\;f_{j}=f(\nu_{j}),\;
\vec a=\left(
\begin{array}{c}
a_{1}\\
...\\
a_{N}
\end{array}
\right)
\end{eqnarray}
The matrix entries $a(\nu_{i},\nu_{j})$ arise from computing integrals (\ref{defa}). In practice these integrals are evaluated on a reference (unit) patch. Such a reference patch together with a vectorial ansatz function is shown in Fig. \ref{fig:fem}(c).

In the above sketch we assumed for simplicity that boundary conditions (\ref{boundary}) are known for the electric field $\MyField E$. However here we want to take the infinite exterior into account. Therefore the eigenvalue problem (\ref{eq:evp1}) has to be solved on an unbounded domain ${\mathbb R}^{2}$. This leads to the computation of leaky modes which enable us to estimate radiation losses. Since our computational domain still has to be of finite size, we apply so-called transparent boundary conditions to $\partial\Omega$. We realize these boundary conditions with the perfectly matched layer (PML) method \cite{BerPML}. Details about our numerical implementation are described in \cite{Zschiedrich2005a}. The propagation constant $k_{z}$ (and the effective refractive index) becomes complex and the corresponding mode is damped according to ${\mathrm{exp}}({-\Im(k_{z})z})$ while propagating along the fiber, see Eq. (\ref{eq:propAnsatz}).

\begin{figure}
(a)\hspace{4.8cm}(b)\hspace{4.8cm}(c)\\
\includegraphics[width=4.5cm,height=4.5cm]{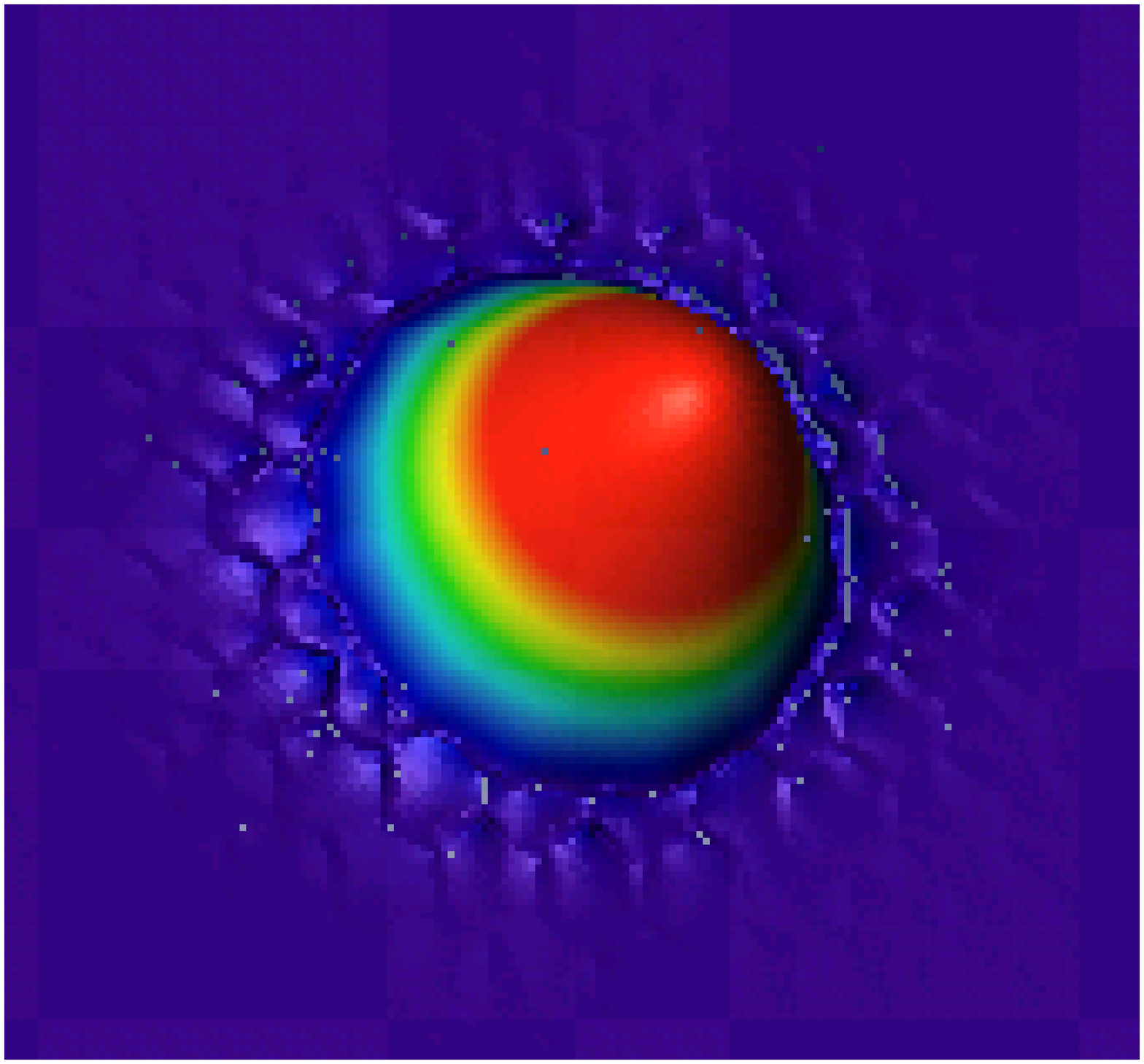}\hfill
\includegraphics[width=4.5cm,height=4.5cm]{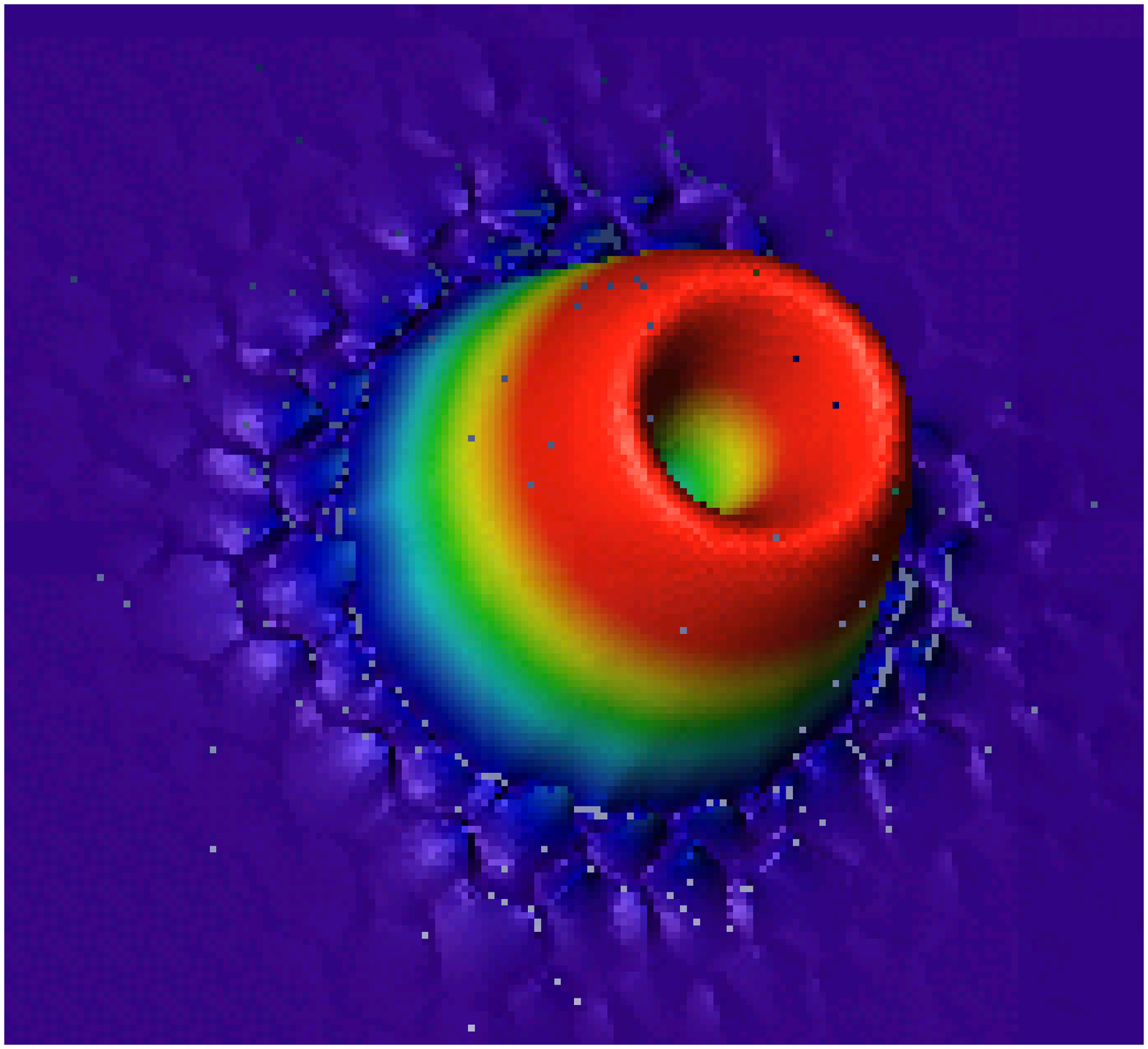}\hfill
\includegraphics[width=4.5cm,height=4.5cm]{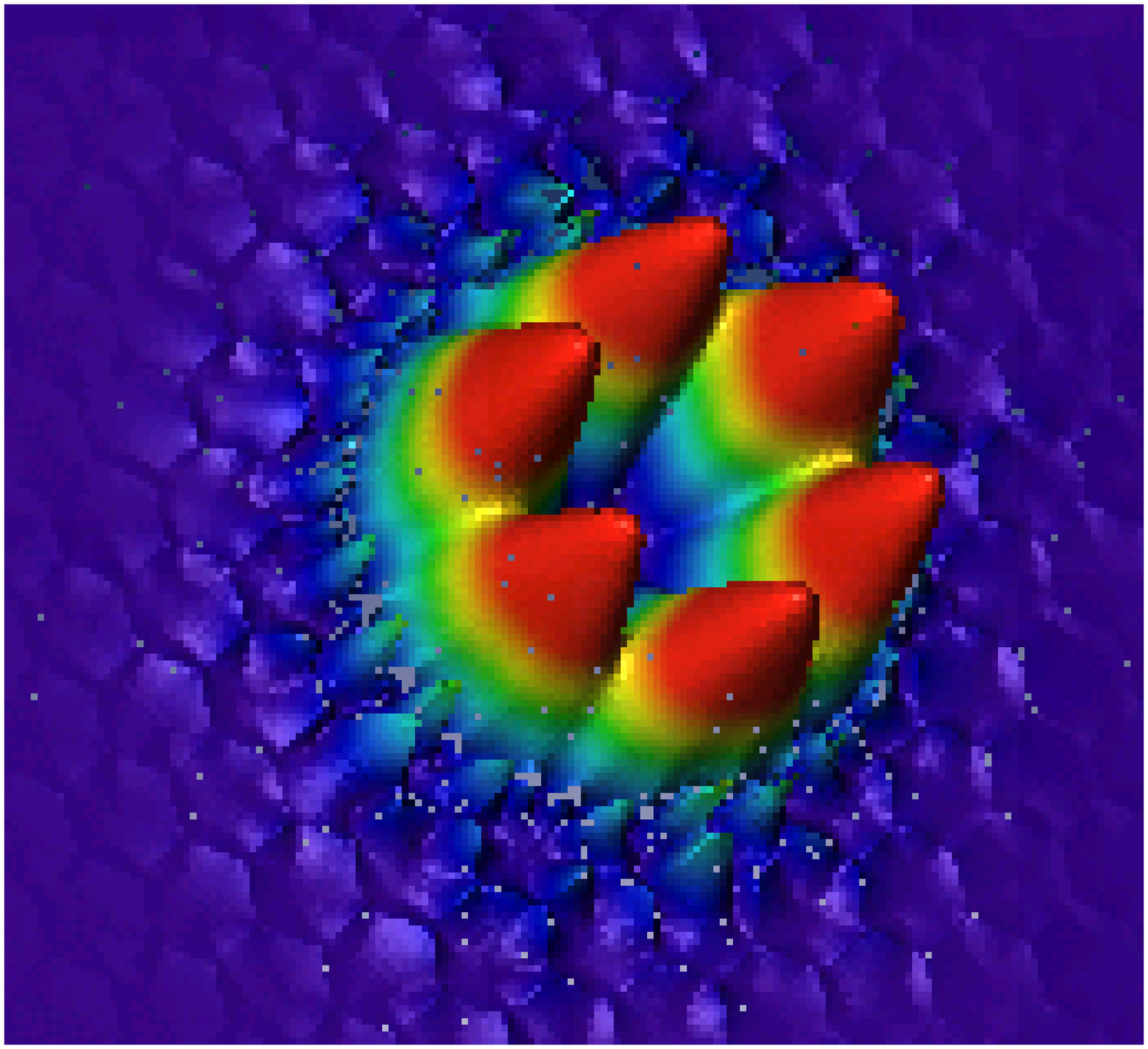}\hfill
\caption{\label{fig:coreModes}First, second and fourth fundamental core modes of HCPCF illustrated in Fig. \ref{fig:hcpcf}(a) - Parameters: $\Lambda=1550\,$nm, $r=300\,$nm, $w=50\,$nm, $t=170\,$nm, $\lambda=589\,$nm, see Fig. \ref{fig:hcpcfTriang}(a) for definition of parameters.}
\end{figure}

Applying the finite element method to propagating mode computation has several advantages \cite{CUC02,BRE00}. The flexibility of triangulations allows the computation of virtually arbitrary structures without simplifications or approximations, as illustrated in Fig. \ref{fig:fem}(b) and \ref{fig:hcpcfTriang}(b). By choosing appropriate ansatz functions $\MyField{\nu}_{i}(x,y)$ for the solution of Maxwell's equations, physical properties of the electric field like discontinuities or singularities can be modeled very accurately and do not give rise to numerical problems. Such discontinuities often appear at glass/air interfaces of photonic crystal fibers, see Fig. \ref{fig:coreModes}. Adaptive mesh-refinement strategies lead to very accurate results and small computational times. Furthermore the FEM approach converges with a fixed convergence rate towards the exact solution of Maxwell-type problems for decreasing mesh width (i.e. increasing number $N$ of sub-patches) of the triangulation. Therefore, it is easy to check if numerical results can be trusted \cite{MON03}.

Especially for complicated geometrical structures the finite element method is better suited for mode computation than other methods. In contrast to the plane wave expansion (PWE) method, whose ansatz functions are spread over the whole computational domain (plane waves) the FEM method uses localized ansatz functions, see Fig. \ref{fig:fem}(c). In order to expand a solution with discontinuities as shown in Fig. \ref{fig:coreModes} a large number of plane waves would be necessary using the PWE method. This leads to slow convergence and large computational times \cite{HOL06}. 

For accurate and fast computation of leaky eigenmodes we have implemented several features into the FEM package JCMsuite. As we will see later e.g. high-order edge elements, a-posteriori error control and adaptive and goal-oriented mesh refinement increase the convergence of numerical results dramatically.

\section{Computation of leaky modes in hollow core photonic crystal fibers}
\begin{figure}[h]
(a)\hspace{9.7cm}(b)\\
\includegraphics[height=4cm]{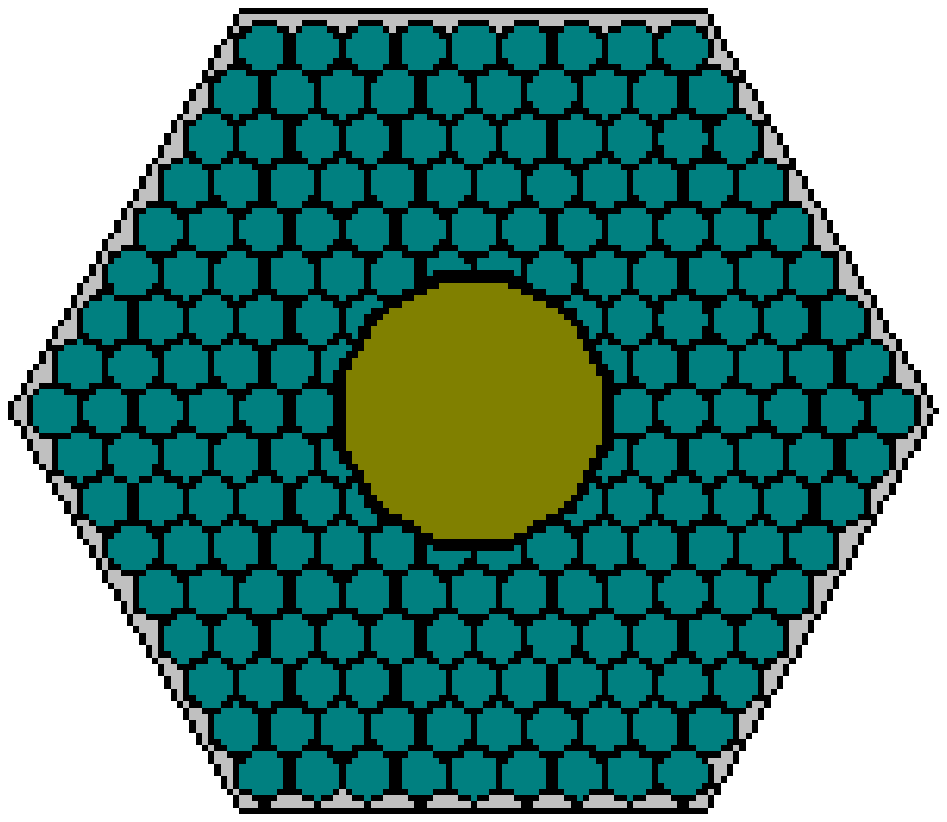}\hfill
\includegraphics[height=4cm]{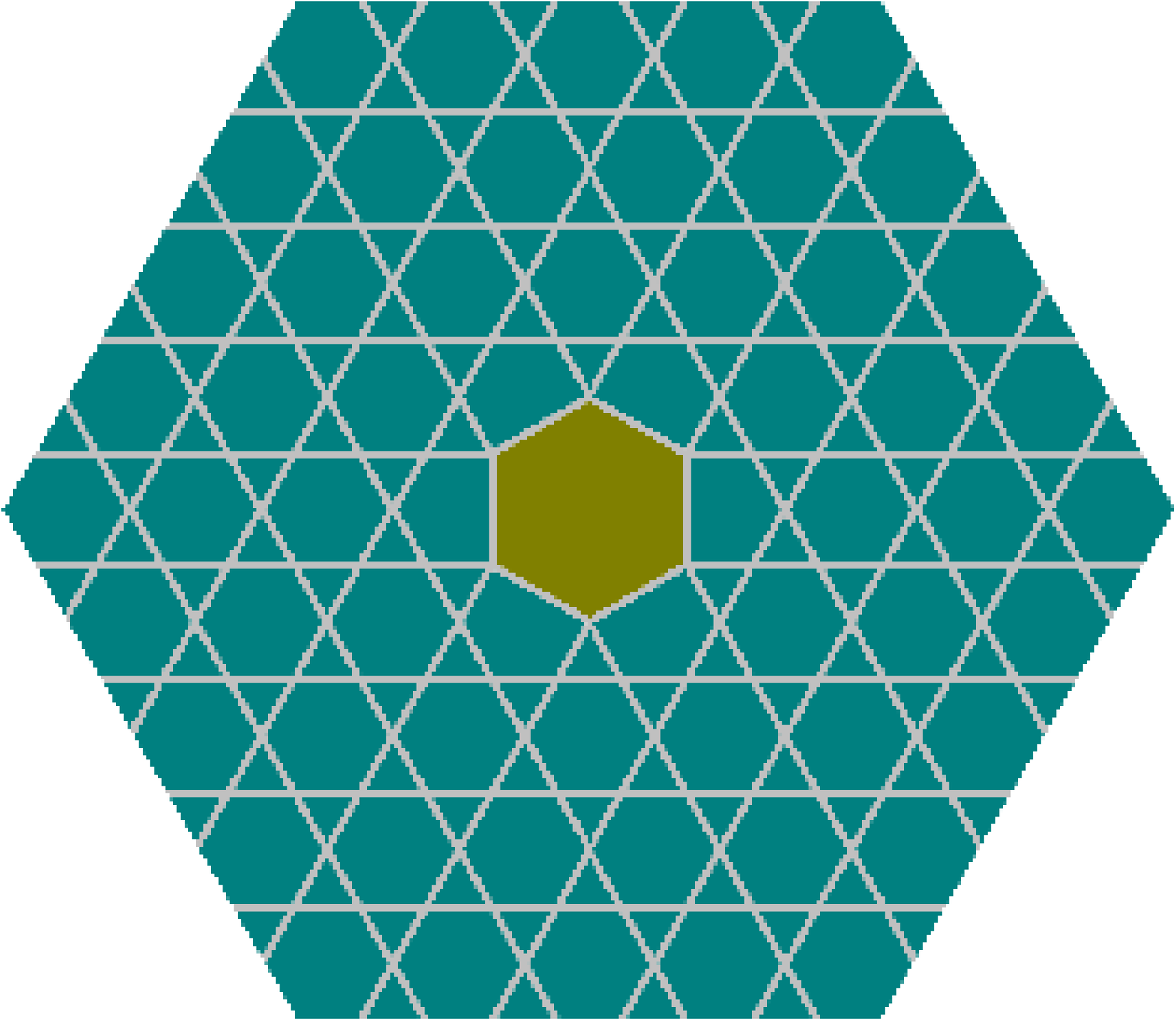}\hfill
\caption{\label{fig:hcpcf}(a) Geometry of 19-cell HCPCF and (b) kagome-structured fiber used for mode computation}
\end{figure}

The only simplification we make for the computation of leaky modes is to extend the fiber cladding to infinity and thereby neglect its finite size. This is justified if the cladding of the fiber is much larger than the microstructured core, and no light entering the cladding is reflected back into the core, which is usually the case. We investigate two different types of hollow-core photonic crystal fibers (HCPCF). The first type has a hollow core which corresponds to 19 omitted hexagonal cladding cells Fig. \ref{fig:hcpcf}(a) the second type is kagome-structured \cite{COU06}, Fig. \ref{fig:hcpcf}(b). 
The cross sections of the HCPCFs depicted in Fig. \ref{fig:hcpcf} have a $\mathrm{C}_{6\mathrm{V}}$ invariance. Therefore it is possible to take only a fraction of the layout as computational domain. At the artificial inner boundaries appropriate boundary conditions have to be stated, setting the tangential magnetic and electric field to $0$ respectively. Table \ref{table:eigenvaluesSymmetry} shows the first and second eigenvalue for the full half and quarter fiber used as computational domain. The geometrical parameters are given in Fig. \ref{fig:coreModes}. The computed values are identical up to the chosen accuracy. The number of unknowns and computational time is of course much smaller for half and quarter fiber cross section. 

The imaginary and real parts of the eigenvalues given in Table \ref{table:eigenvaluesSymmetry} differ by up to 11 orders of magnitude. The complex eigenvalue leads to a dampening of the mode according to 
\begin{equation}
  \label{eq:dampk_z}
  |\MyField{E}|^{2}\propto e^{-2\Im(k_{z})z},
\end{equation}
see Eq. (\ref{eq:propAnsatz}). Computing the fiber without transparent boundary conditions but setting the tangential component of the electric field to $0$ at the outer boundary we get $n^{1}_{\mathrm{eff}} = 0.99826580015$ for the first eigenvalue, i.e. a real eigenvalue. Comparing to Table \ref{table:eigenvaluesSymmetry} we see that taking into account the finite size of the fiber and coupling to the exterior does not change the real part. However when analyzing radiation losses of a fiber, the imaginary part of the effective refractive index $n_{\mathrm{eff}}$ is the quantity of interest. Computation of leaky modes with only small losses is therefore a multi-scale problem which is numerically very difficult to handle. In the next section we will introduce a goal-oriented error estimator which controls the grid refinement in order to obtain an accurate imaginary part as fast as possible.
\begin{table}
\centering
\begin{tabular}{lcccc}
&unknowns & 1st eigenvalue & 2nd eigenvalue  \\
\hline
full fiber   &3070641    &  $0.99826580337 + 8.9297 \cdot 10^{-12}i$ &$0.992141777 + 1.3312\cdot 10^{-10}i$  \\
half fiber   & 1567377   & $0.99826580252 + 8.9311 \cdot 10^{-12}i$ &$0.992141758 + 1.3314\cdot 10^{-10}i$ \\
quarter fiber &781936 & $0.99826580254 + 8.9311 \cdot 10^{-12}i$ &$0.992141754 + 1.3315\cdot 10^{-10}i $\\
\end{tabular}
\caption{\label{table:eigenvaluesSymmetry}First, second and third eigenvalue computed with full, half and quarter fiber as computational domain.}
\end{table}
\section{Goal oriented error estimator}
The finite element method enables us to refine patches of the unstructured grid only locally. We use error estimators to control the refinement process of the grid. Usually such an error estimator is defined by minimizing a target functional $j(\MyField{E})$, i.e. it is \textit{goal-oriented}. The target functional depends on the solution of the electric field $\MyField{E}$. Since we are interested in radiation losses this will be the imaginary part of the propagation constant \cite{ZSC07}. From Maxwell's equations applied to a waveguide structure one can derive the following expression for the imaginary part of the propagation constant:
\begin{eqnarray}
  \label{eq:kzDef}
  \Im \left( k_{z} \right) = \frac{P_{\partial \Omega}(\MyField{E})}{2 P_{\Omega}(\MyField{E})},
\end{eqnarray}
where
\begin{eqnarray}
  \label{eq:powerFluxes}
  P_{\Omega}(\MyField{E}) & = & \frac{1}{2\Re (\omega)} \Im\left( \int_{\Omega} \left[  \bar{\MyField{E}} \times \frac{1}{\MyTensor{\mu}}  \nabla_{k_z} \MyField{E} \right]\cdot \hat{n}_{z} \right) \\
P_{\partial \Omega}(\MyField{E}) & = & \frac{1}{2 \Re(\omega)} \Im\left(\int_{\partial \Omega}  \left[  \bar{\MyField{E}} \times \frac{1}{\MyTensor{\mu}}  \nabla_{k_z} \MyField{E} \right]\cdot \hat{n}_{xy} \right) 
\end{eqnarray}
is the power flux of the electric field through the cross section $\Omega$ of the computational domain and the in plane power flux across the boundary $\partial \Omega$ of the computational domain respectively. Equation (\ref{eq:kzDef}) also shows us that the imaginary part of $k_{z}$ reflects radiation leakage from the photonic crystal fiber to the exterior.
The right hand side of (\ref{eq:kzDef}) is the target nonlinear functional $j(\MyField{E})$ which is used for the control of the error estimator. The finite element mesh should now be adapted such that $j(\MyField{E}_{h})-j(\MyField{E})$ is minimized, where $\MyField{E}_{h}$ is the finite element solution and $\MyField{E}$ the exact solution. This task can be embedded into the framework of optimal control theory \cite{Becker:01a}. We define the trivial optimization problem:
\begin{eqnarray}
  \label{eq:opt}
  j(\MyField{E})-j(\MyField{E}_{h}) = \min_{\MyField{\Psi} \in\ H(curl)} \left\{j(\MyField{\Psi})-j(\MyField{E}_{h}) \; : \; \; a\left(\MyField{\Phi}, \MyField{\Psi} \right) = f(\MyField{\Phi})  \;  \forall \MyField{\Phi} \in H(curl) \right\}.
\end{eqnarray}
This formulation is trivial because the restriction simply states that $\Psi$ is a solution of Maxwell's equation. The minima of (\ref{eq:opt}) correspond to stationary points of the Lagrangian density
\begin{eqnarray*}
\mathcal{L}(\MyField{E}, \MyField{E^{*}}) = j(\MyField{E})-j(\MyField{E}_{h}) + f(\MyField{E}^{*})- a\left(\MyField{\MyField{E}^{*}}, \MyField{E} \right),
\end{eqnarray*}     
where $\MyField{E}^{*}$ denotes the 'dual' variable (Lagrangian multiplier). Hence, we seek the solution $(\MyField{E}, \MyField{E}^{*})$ to the Euler-Lagrange system
\begin{eqnarray}
a\left(\MyField{\Phi}, \MyField{E} \right) & = & f(\MyField{\Phi})  \qquad \forall \MyField{\Phi} \in H(curl) \label{Eqn:EulerLagrangePrimary} \\
a\left(\MyField{E}^{*}, \MyField{\Phi} \right) & = & j'(\MyField{E};    \MyField{\Phi})  \qquad \forall \MyField{\Phi} \in H(curl). \label{Eqn:EulerLagrangeDual}
\end{eqnarray}
Equation (\ref{Eqn:EulerLagrangePrimary}) is the variational form of the original Maxwell's equations. In the dual Eq. (\ref{Eqn:EulerLagrangeDual}) the target functional $j(\MyField{E})=k_{z}$ appears in form of its linearization $j'(\MyField{E}).$  A finite element discretization of the Euler-Lagrange system yields a supplemental discrete problem
\begin{eqnarray*}
a\left(\MyField{E}_{h}^{*}, \MyField{\Phi}_h \right) & = & j'(\MyField{E}_{h};    \MyField{\Phi}_{h})  \quad \forall \MyField{\Phi}_h   \in V_{h}. 
\end{eqnarray*}
To quantify the error of the finite element solution $\MyField{E}_{h}$ we introduce the primal and dual residuals: 
\begin{eqnarray}
\rho(\MyField{E}_{h}; \cdot) & =  & f(\cdot)-a\left(\cdot, \MyField{E}_{h} \right)\label{residuumsa}\\
\rho(\MyField{E}^{*}_{h}; \cdot) & =  & j'(\MyField{E}_{h}; \cdot)-a\left(\MyField{E}^{*}_{h}, \cdot \right)\label{residuumsb}.
\end{eqnarray}
The residuals quantify the error inserting the approximate solution into the exact non-discretized Maxwell's equations. For the exact solution one finds $\rho(\MyField{E}_{h};\MyField {E})=0$ and $\rho(\MyField{E}_{h}^{*};\MyField{E})=0$. These residuals are computed for each patch and only patches with the largest residuals are refined. More details about the mathematical formulation and implementation can be found in \cite{ZSC07}. A closer look to $\rho(\MyField{E}_{h}; \cdot)$ which is also used for field-energy based adaptive refinement will be given in the next section.
\begin{figure}[]
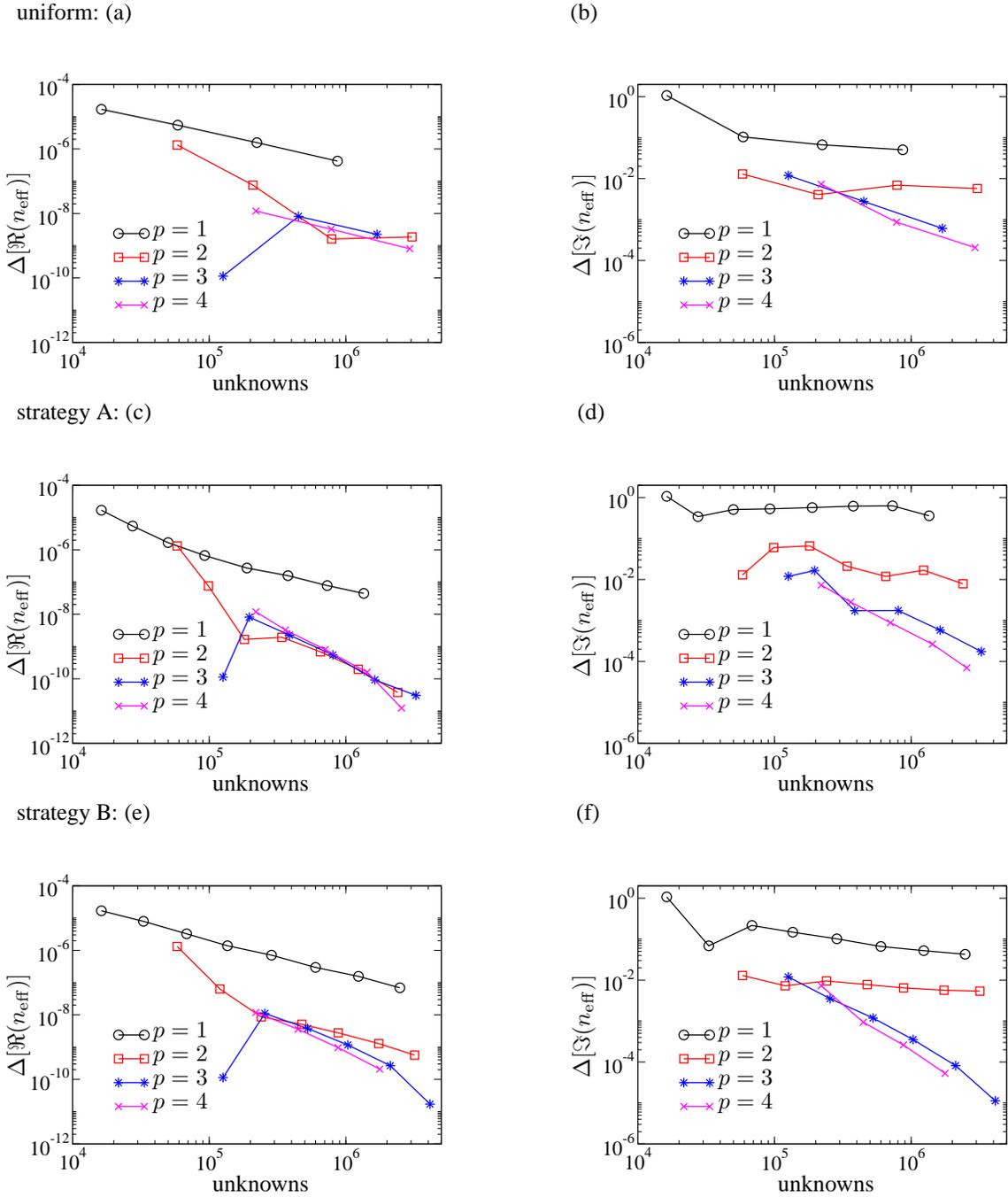

\psfrag{deltare}{$\Delta[\Re(n_{\mathrm{eff}})]$}
\psfrag{deltaim}{$\Delta[\Im(n_{\mathrm{eff}})]$}
\psfrag{unk}{unknowns}
\psfrag{p=1}{$p=1$}
\psfrag{p=2}{$p=2$}
\psfrag{p=3}{$p=3$}
\psfrag{p=4}{$p=4$}
\psfrag{no}{uniform}
\psfrag{go}{strategy B}
\psfrag{ad}{strategy A}
uniform: (a) \hspace{6.5cm}(b)\vspace{7mm}\\
\includegraphics[width=6.4cm]{fig/convSFBrealNo.eps}\hfill
\includegraphics[width=6.4cm]{fig/convSFBimagNo.eps}\hfill\\
strategy A: (c)\hspace{6.4cm}(d)\vspace{7mm}\\
\includegraphics[width=6.4cm]{fig/convSFBrealAd.eps}\hfill
\includegraphics[width=6.4cm]{fig/convSFBimagAd.eps}\hfill\\
strategy B: (e)\hspace{6.4cm}(f)\vspace{7mm}\\
\includegraphics[width=6.4cm]{fig/convSFBrealGo.eps}\hfill
\includegraphics[width=6.4cm]{fig/convSFBimagGo.eps}\hfill\\
\caption{\label{fig:hcpcfConv}Relative error of fundamental eigenvalue in dependence on number of unknowns of FEM computation for different refinement strategies and finite element degrees $p$. Parameters: $\Lambda=1550\,$nm, $r=300\,$nm, $w=50\,$nm, $t=170\,$nm, 6 cladding rings, wavelength $\lambda=589\,$nm. Adaptive refinement strategy A minimizes residuum (\ref{residuumsa}), strategy B minimizes residua (\ref{residuumsa}) and (\ref{residuumsb})}
\end{figure}
\begin{figure}
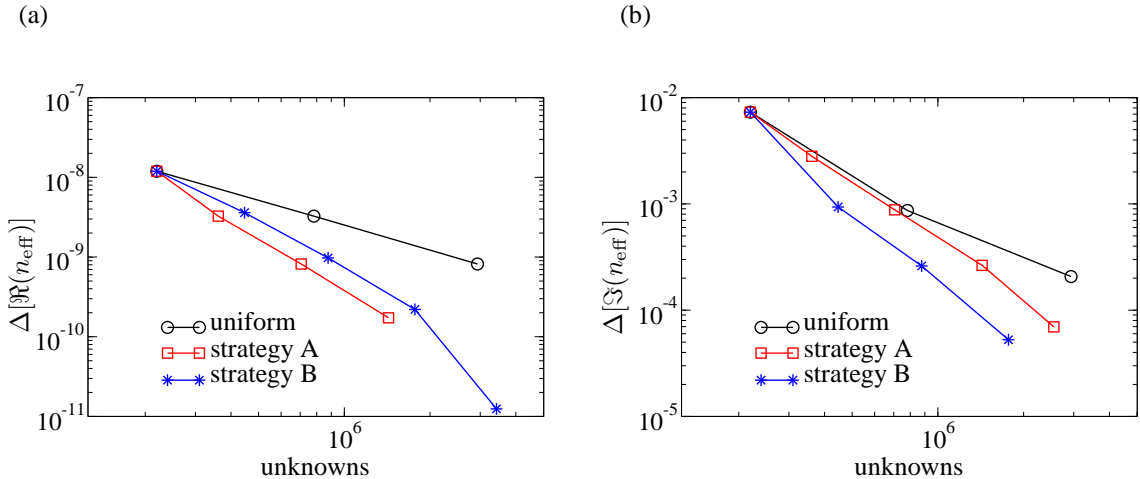

\psfrag{deltare}{$\Delta[\Re(n_{\mathrm{eff}})]$}
\psfrag{deltaim}{$\Delta[\Im(n_{\mathrm{eff}})]$}
\psfrag{unk}{unknowns}
\psfrag{p=1}{$p=1$}
\psfrag{p=2}{$p=2$}
\psfrag{p=3}{$p=3$}
\psfrag{p=4}{$p=4$}
\psfrag{no}{uniform}
\psfrag{go}{strategy B}
\psfrag{ad}{strategy A}
(a)\hspace{7.6cm}(b)\vspace{7mm}\\
\includegraphics[width=7cm]{fig/convFEM4NoAdGoReal.eps}\hfill
\includegraphics[width=7cm]{fig/convFEM4NoAdGoImag.eps}\hfill
\caption{\label{fig:hcpcfConv2}Relative error of fundamental eigenvalue in dependence on number of unknowns of FEM computation for different refinement strategies and finite element degree $p=3$. Parameters: $\Lambda=1550\,$nm, $r=300\,$nm, $w=50\,$nm, $t=170\,$nm, 6 cladding rings, wavelength $\lambda=589\,$nm. (a) uniform refinement, (b) adaptive refinement, (c) goal-oriented refinement}
\end{figure}
\section{Convergence of eigenvalues using different error estimators}
In this section we will look at the convergence of the computed eigenvalues. As computational domain we use the HCPCF depicted in Fig. \ref{fig:hcpcf}(a). We compare three different refinement strategies namely uniform refinement and two different adaptive goal-oriented refinement strategies. In an uniform refinement step each triangle is subdivided into 4 smaller ones. In an adaptive refinement step a predefined resdiuum $\rho$ of the numerical solution is minimized by refining only triangles with largest values for this residuum. The two different adaptive strategies we are using differ by the chosen residua. The first one only uses (\ref{residuumsa}). We will refer to it as \textit{strategy A}. The second \textit{strategy B} which was introduced in the previous section uses (\ref{residuumsb}) in addition to (\ref{residuumsa}).

Let us have a closer look at the common residuum (\ref{residuumsa}) of both strategies. We start with the weak formulation of Maxwell's equations (\ref{eq:MWweak2}) on a sub patch $\Omega_{i}$ of our triangulation and reverse the partial integration:
\begin{eqnarray}
  \label{eq:partialReverse}
0= \int_{\Omega_{i}}\overline{\MyField{\Phi}}\cdot\left[\nabla_{k_z}\times\frac{1}{\MyTensor{\mu}}\nabla_{k_z}\times\MyField{ E}-\frac{\omega^{2}\MyTensor{\epsilon}}{c^{2}}\MyField{ E}\right]d^{3}r-\int_{\Gamma_{i}}\overline{\MyField{\Phi}}\cdot\left[\left(\left[\left[\frac{1}{\MyTensor{\mu}}\nabla_{k_z}\times\MyField{E}\right]\right]\right)\times\vec n\right]d^{2}r,
\end{eqnarray}
where $\Gamma_{i}$ is the boundary of the sub patch $\Omega_{i}$ and $\left[\left[\frac{1}{\MyTensor{\mu}}\nabla_{k_z}\times\MyField{E}\right]\right]$ is the difference of the electric field $\MyField{E}$ and the permeability $\MyTensor{\mu}$ on both sides of this boundary. Since Eq. (\ref{eq:partialReverse}) holds for arbitrary $\MyField{\Phi}$ the terms in brackets in both integrals have to vanish. For an exact solution to Maxwell's equations the first term vanishes because this is the Maxwell equation itself and the second term because the tangential component of $\frac{1}{\MyTensor{\mu}}\nabla_{k_z}\times\MyField{E}$  is continuous across a boundary. Since we approximate the exact solution by $\MyField{E}_{h}$ both terms will generally not vanish. Therefore we define the residuum $\rho_{i}$ of sub patch $\Omega_{i}$:
\begin{equation}
  \label{eq:residuum}
\rho_{i}(\MyField{E}_{h};\MyField{E}_{h})=h_{i}^{2}\int_{\Omega_{i}}\left|\nabla_{k_z}\times\frac{1}{\MyTensor{\mu}}\nabla_{k_z}\times\MyField{ E}_{h}-\frac{\omega^{2}\MyTensor{\epsilon}}{c^{2}}\MyField{ E}_{h}\right|^{2}d^{3}r+h_{i}\int_{\Gamma_{i}}\left|\left(\left[\left[\frac{1}{\MyTensor{\mu}}\nabla_{k_z}\times\MyField{E}_{h}\right]\right]\right)\times\vec n\right|^{2}d^{2}r,
\end{equation}
where $h_{i}$ is the size of the sub patch. For the solution $\MyField{E}_{h}=\MyField{E}$ of Maxwell's equations we find $\rho_{i}(\MyField{E}_{h};\MyField{E})=0$. The residuum is therefore a measure how well the approximation $\MyField{E}_{h}$ fulfills the exact Maxwell's equations.

Figure \ref{fig:hcpcfConv} shows the relative error of the fundamental eigenvalue $n_{\mathrm{eff}}^{1}$ in dependence on the number of unknowns of the FEM computation for different refinement strategies and finite element degrees $p$. The real part of the effective refractive index converges for all finite element degrees and all refinement strategies, see Fig. \ref{fig:hcpcfConv} (a),(c),(e). For higher finite element degrees we find faster convergence. For the lowest finite element order and coarsest grid (with $\approx 16000$ unknowns) we have a relative error of $\approx 10^{-5}$. This error decreases down to $10^{-11}$ for fourth order elements with $4\cdot 10^{6}$ unknowns. Figure \ref{fig:hcpcfConv} (b),(d),(f) show that the imaginary part converges much slower. For uniform refinement and finite element degrees of $p=1,2$ we do not find convergence at all. The adaptive strategies A and B refinement also lead to poor convergence for low finite element degrees. For an accurate computation of small losses therefore high finite element degrees are necessary. For a relative error of $10^{-3}=0.1\%$ we already need $\approx10^{6}$ unknowns for $p=3,4$. In Fig. \ref{fig:hcpcfConv2}(a) and (b) the different refinement strategies are compared for fixed finite element order $p=4$. For the real part we find fastest convergence for adaptive strategy A. As explained before the corresponding error estimator refines those triangles where the electric field has a large deviation from the exact solution. The error estimator of strategy B also uses this residuum but furthermore the residuum of the dual problem, Eq. (\ref{residuumsa}) and (\ref{residuumsb}). Therefore it also converges faster than uniform refinement but slower than strategy A. For the imaginary part the adaptive refinement strategy B shows fastest convergence. Strategy A shows almost no benefit for the first two refinement steps in comparison to uniform refinement. For the exact result for the convergence plots we used the most accurate result ($p=4$) for the real and imaginary part obtained from strategy A and B respectively.

\begin{figure}[]
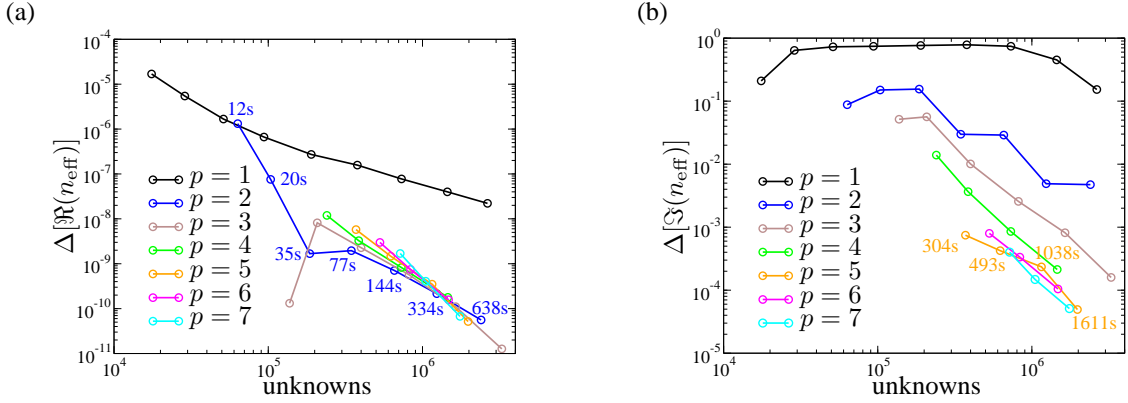

\psfrag{deltare}{$\Delta[\Re(n_{\mathrm{eff}})]$}
\psfrag{deltaim}{$\Delta[\Im(n_{\mathrm{eff}})]$}
\psfrag{unk}{unknowns}
\psfrag{p=1}{$p=1$}
\psfrag{p=2}{$p=2$}
\psfrag{p=3}{$p=3$}
\psfrag{p=4}{$p=4$}
\psfrag{p=5}{$p=5$}
\psfrag{p=6}{$p=6$}
\psfrag{p=7}{$p=7$}
\psfrag{no}{uniform}
\psfrag{go}{goal-oriented}
\psfrag{ad}{adaptive}
(a)\hspace{8cm}(b)\vspace{0mm}\\
\phantom{........}
\includegraphics[width=6cm]{fig/convRealRGB.eps}\hfill
\includegraphics[width=6cm]{fig/convImagRGB.eps}\hfill
\caption{\label{fig:hcpcfConvAda}Relative error of fundamental eigenvalue in dependence on number of unknowns of FEM for adaptive refinement and different finite element degrees $p$. Parameters: $\Lambda=1550\,$nm, $r=300\,$nm, $w=50\,$nm, $t=170\,$nm, 6 cladding rings, wavelength $\lambda=600\,$nm. Computational times are shown.}
\end{figure}

Fig. \ref{fig:hcpcfConvAda} shows the benefit of high order finite elements. Here the adaptive refinement strategy A was used corresponding to Fig. \ref{fig:hcpcfConv}(c),(d) but with finite element degree up to order $p=7$. While the real part of the eigenvalue converges very fast even with order $p=2$ the imaginary part can be computed much more accurate with higher order finite elements. Compared to $p=2$ using orders greater than $p=4$ the relative error is 2 orders of magnitude smaller for the same number of unknowns. Computational times for an $2.6\,$GHz AMD Opteron processor system are also shown. 

A comparison of the convergence and computation efficiency between our finite element package and the MIT Photonic-Bands (MPB) package (plane wave expansion method) is presented in \cite{Maerz2004a} where bloch modes of photonic crystal structures were computed. The FEM computations showed a much higher convergence rate. Results of the same accuracy could be computed over 100 times faster then with the MPB package. In \cite{HOL06} FEM computations of guided modes in HCPCFs are also compared to plane wave expansion (PWE) computations. Eigenmodes which were computed in about a minute with our finite element package took several hours with the PWE method.

\section{Optimization of HCPCF design}
\begin{figure}[ht]
\psfrag{neffimag}{$\Im(n_{\mathrm{eff}})$}
\psfrag{lambda}{$\Lambda$}
\psfrag{wlabel}{$w$}
\psfrag{core}{$t$}
\psfrag{nrows}{cladding rings}
\psfrag{hole}{$r$}
(a)\hspace{8.5cm}(b)\hspace{5.5cm}\\
\includegraphics[width=6cm]{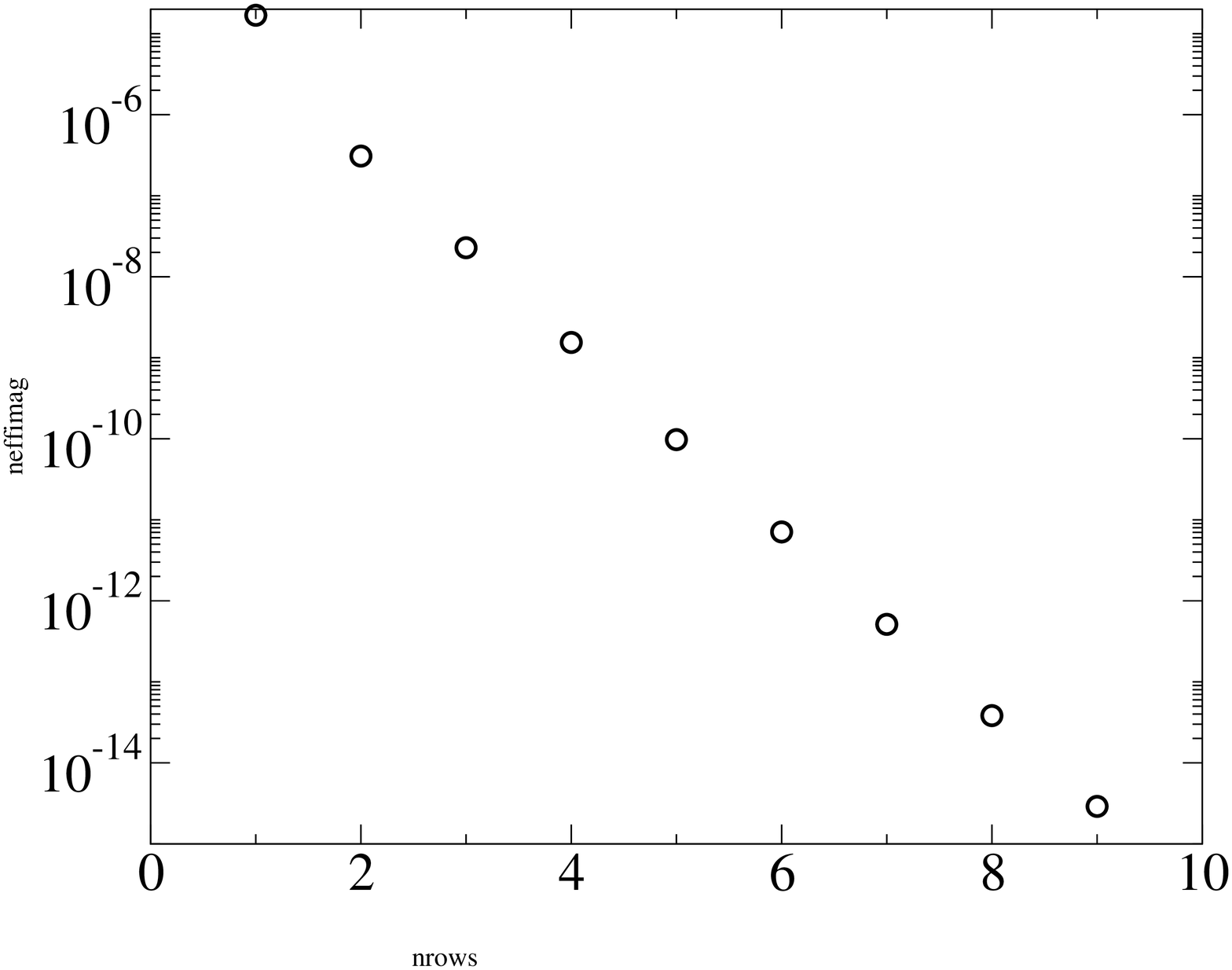}\hfill
\includegraphics[width=6cm]{fig/pitchScan2_.eps}\hfill\\
(c)\hspace{8.5cm}(d)\hspace{5.5cm}\\
\includegraphics[width=6cm]{fig/coreSurroundScan_.eps}\hfill
\includegraphics[width=6cm]{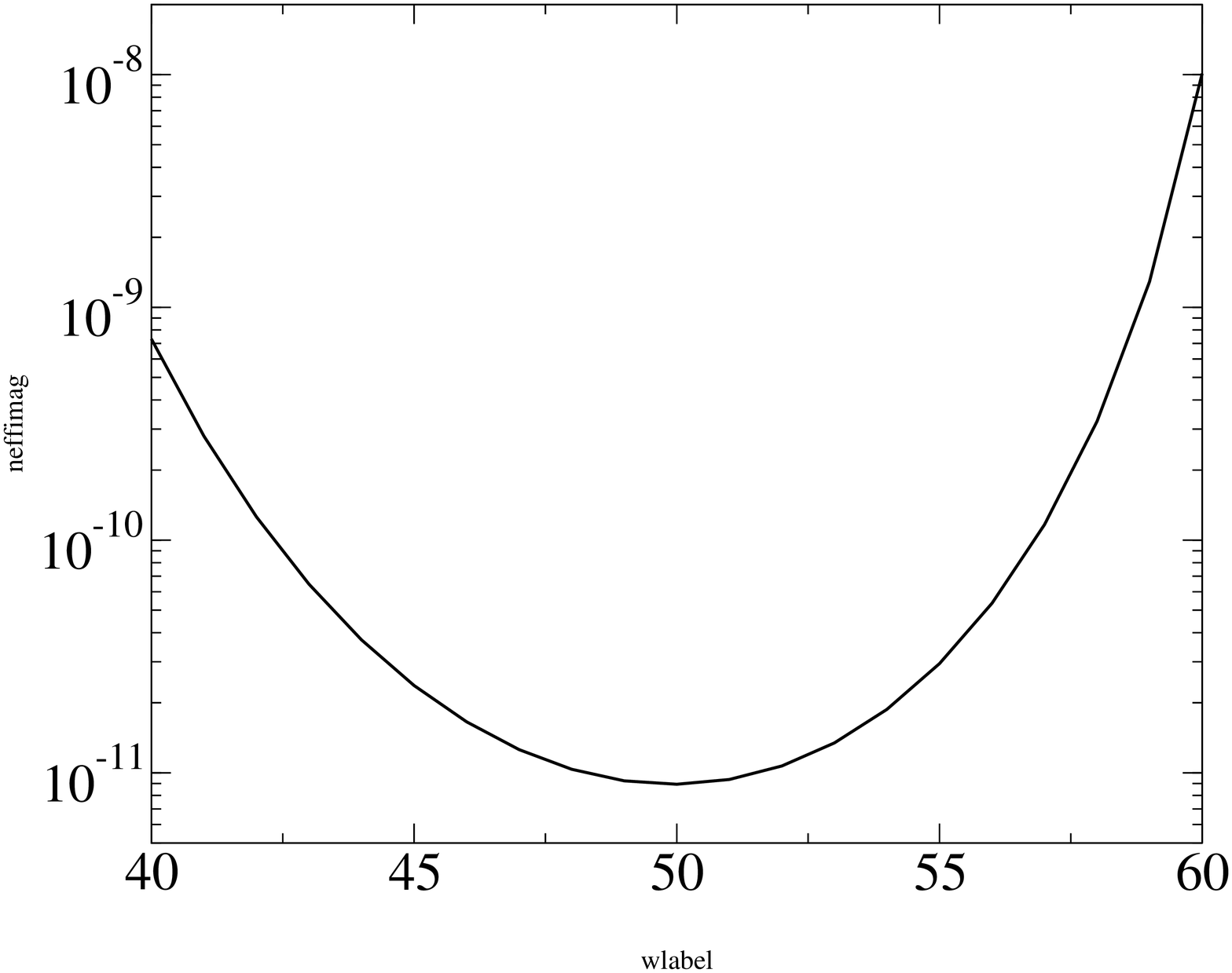}\hfill\\
(e)\\
\includegraphics[width=6cm]{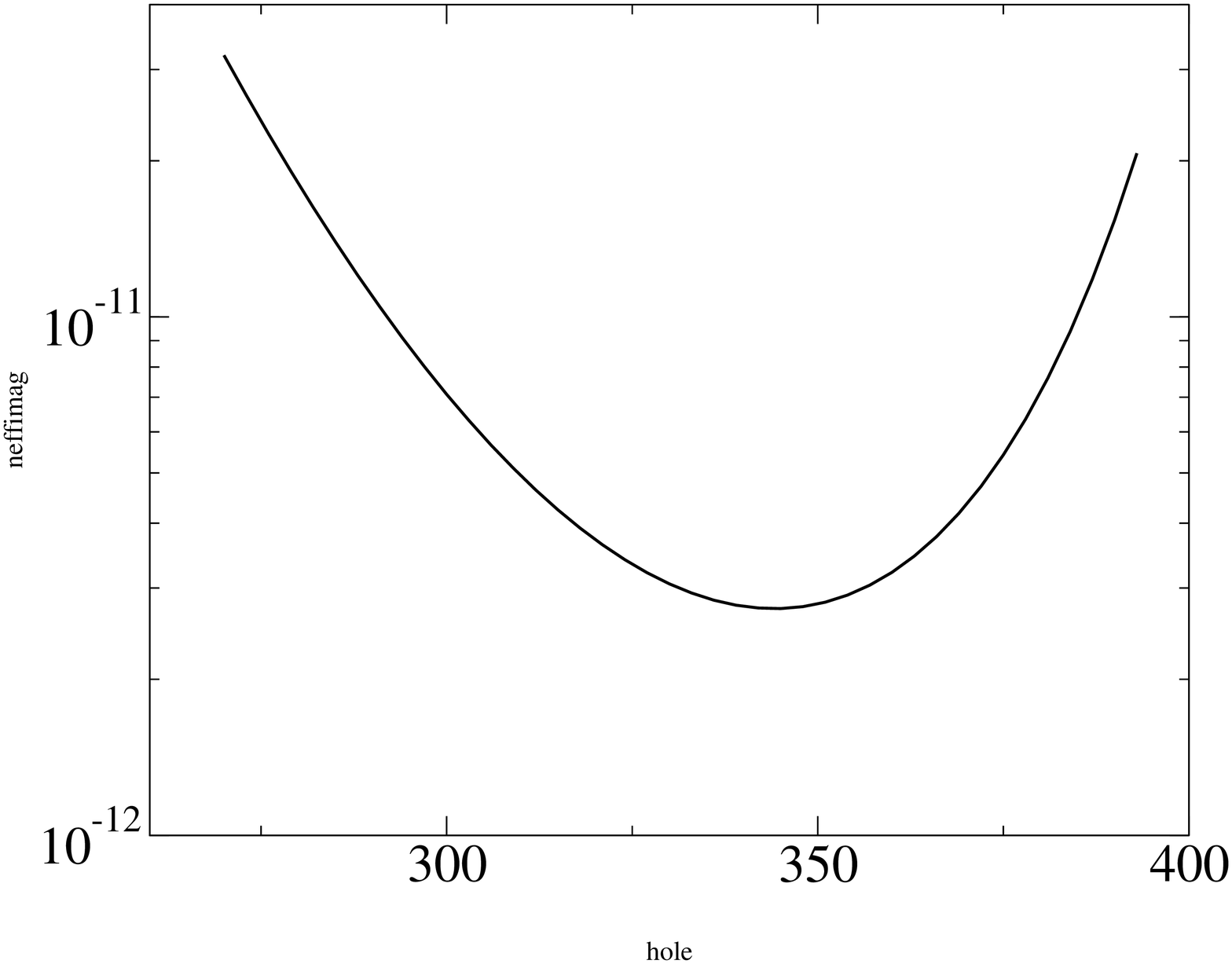}
\caption{\label{fig:geoScan}Imaginary part of effective refractive index $\Im(n_{\mathrm{eff}})$ in dependence on: (a) number of cladding rings, (b) pitch $\Lambda$, (c) core surround thickness $t$, (d) strut thickness $w$, (e) hole edge radius $r$. Parameters: $\Lambda=1550\,$nm, $r=300\,$nm, $w=50\,$nm, $t=170\,$nm, 6 cladding rings, wavelength $\lambda=589\,$nm.}
\end{figure}

Since we are enabled to compute radiation losses very accurately we can use our method to optimize the design of photonic crystal fibers to reduce radiation losses.
The basic fiber layout is a 19-cell core with rings of hexagonal cladding cells, Fig. \ref{fig:hcpcf}(a). Since these cladding rings prevent leakage of radiation to the exterior for core guided modes we expect that an increasing number of cladding rings reduces radiation leakage and therefore reduces $\Im(n_{\mathrm{eff}})$. This is confirmed by our numerical simulations shown in Fig. \ref{fig:geoScan}. The radiation leakage decreases exponentially with the number of cladding rings and thereby the thickness of the photonic crystal structure. This behavior agrees with the exponential dampening of light propagating through a photonic crystal structure with frequency in the photonic band gap.
\begin{figure}[ht]
(a)\hspace{8.5cm}(b)\\
\includegraphics[width=6cm]{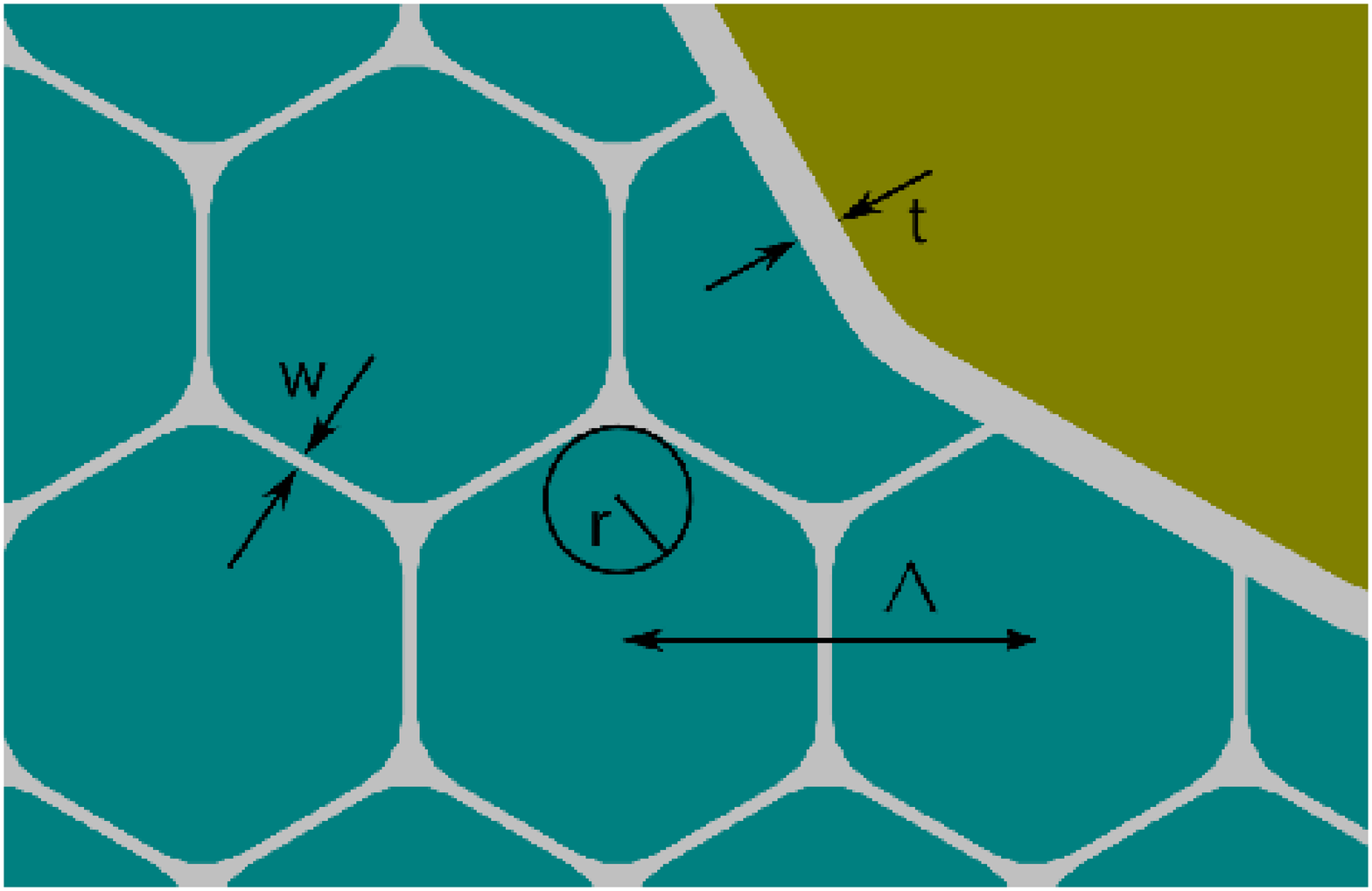}\hfill
\includegraphics[width=6cm]{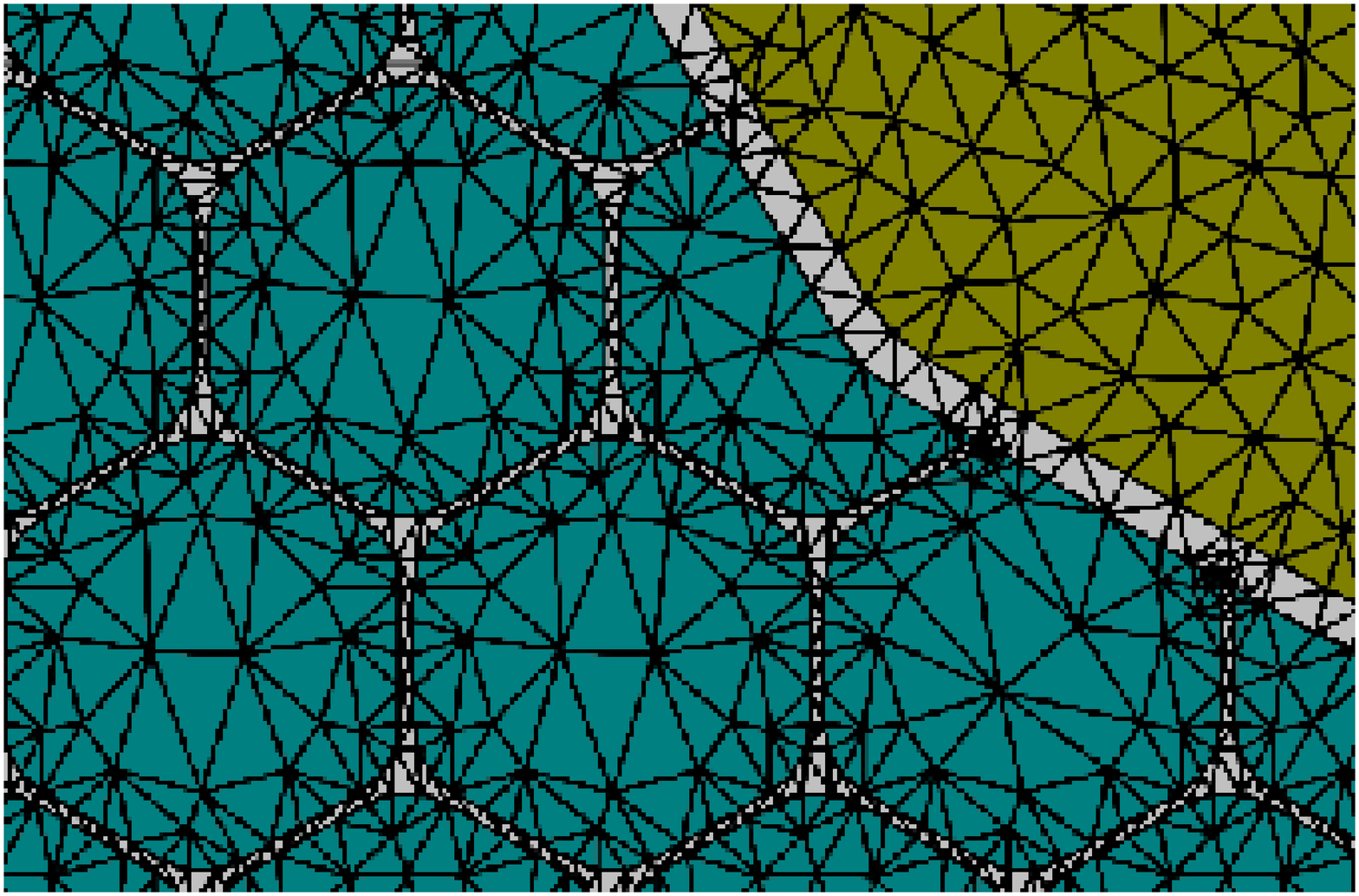}
\caption{\label{fig:hcpcfTriang}(a) geometrical parameters describing HCPCF: pitch $\Lambda$, hole edge radius $r$, strut thickness $w$, core surround thickness $t$; (b) detail from a triangulation of HCPCF. Due to the flexibility of triangulations all geometrical features of the HCPCF are resolved.}
\end{figure}
For our further analysis we fix the number of cladding rings to 6. The free geometrical parameters are the pitch $\Lambda$, hole edge radius $r$, strut thickness $w$, and core surround thickness $t$ depicted in Fig. \ref{fig:hcpcfTriang}(a) together with the triangulation \ref{fig:hcpcfTriang}(b). Fig. \ref{fig:geoScan} shows the imaginary part of the effective refractive index in dependence on these parameters. For each scan all but one parameter were fixed. For the strut thickness $w$ and the hole edge radius $r$ we find well-defined optimal values which minimize $\Im(n_{\mathrm{eff}})$. For pitch $\Lambda$ and core surround thickness $t$ a large number of local minima and maxima can be seen. Now we want to optimize the fiber design using multidimensional optimization with the Nelder-Mead simplex method\cite{HOL06}. To reduce the number of optimization parameters we fix the hole edge radius to the determined minimum at $r=354\,$nm since its variation has the weakest effect on $\Im(n_{\mathrm{eff}})$. For optimization we have to choose starting values for $\Lambda$, $t$ and $w$. Since the simplex method searches for local minima we have to decide in which local minimum of $\Lambda$ and $t$ we want to search. We choose $t=152\,$nm since here $\Im(n_{\mathrm{eff}})$ has a global minimum and $\Lambda=1550\,$nm since the bandwidth of this minimum is much larger than for the global minimum at $\Lambda=1700\,$nm.
Optimization yields a minimum value of 
$\Im(n_{\mathrm{eff}})=5\cdot 10^{-15}\frac{1}{\mathrm{m}}$ for the imaginary part of the effective refractive index. The corresponding geometrical parameters are $\Lambda=1597\,$nm, $w=38\,$nm , $t=151\,$nm.

\section{Kagome-structured fibers}
\begin{figure}[ht]
(a)\hspace{4.7cm}(b)\hspace{4.7cm}(c)\\
\includegraphics[height=4.1cm]{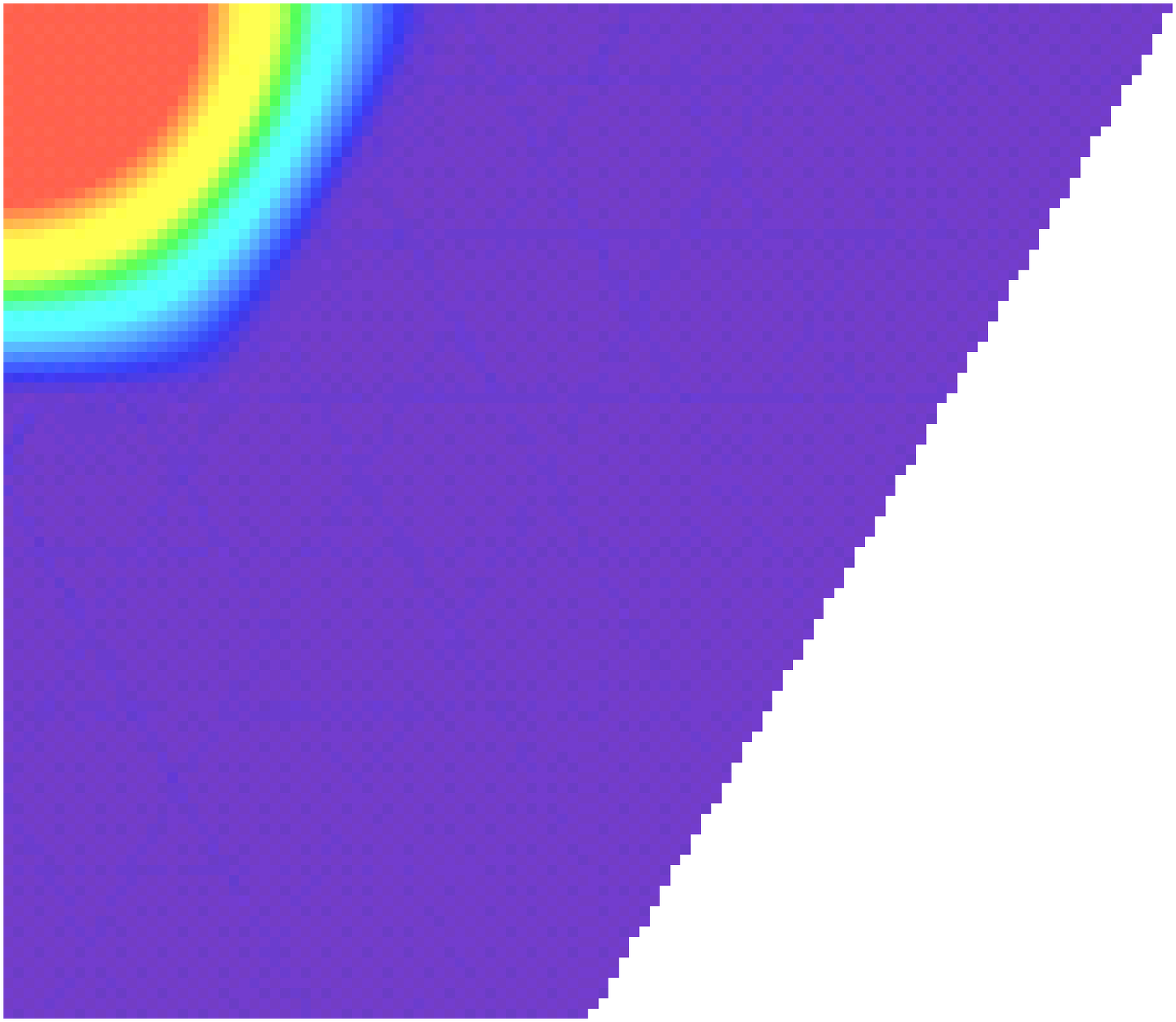}\hfill
\includegraphics[height=4.1cm]{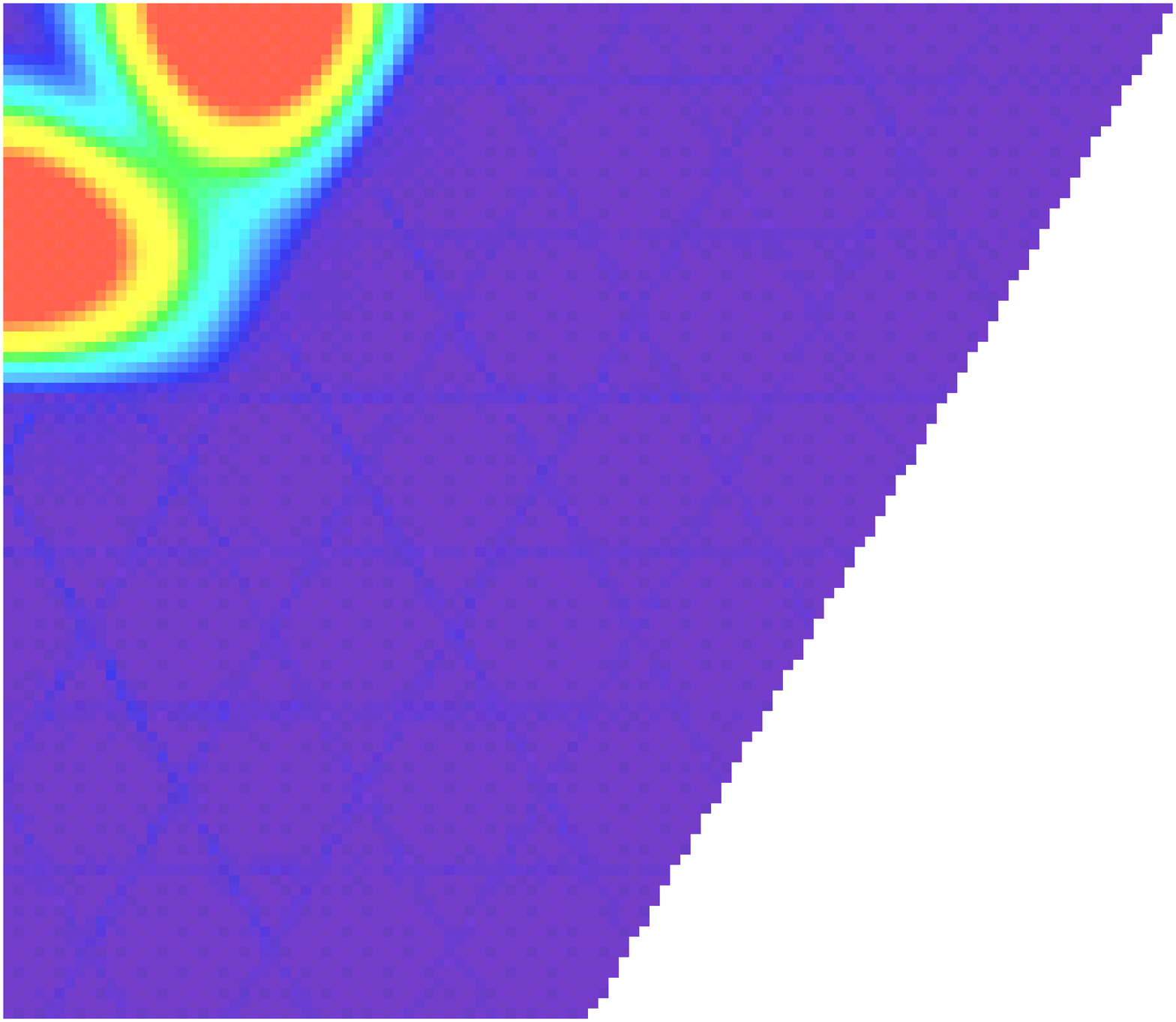}\hfill
\includegraphics[height=4.1cm]{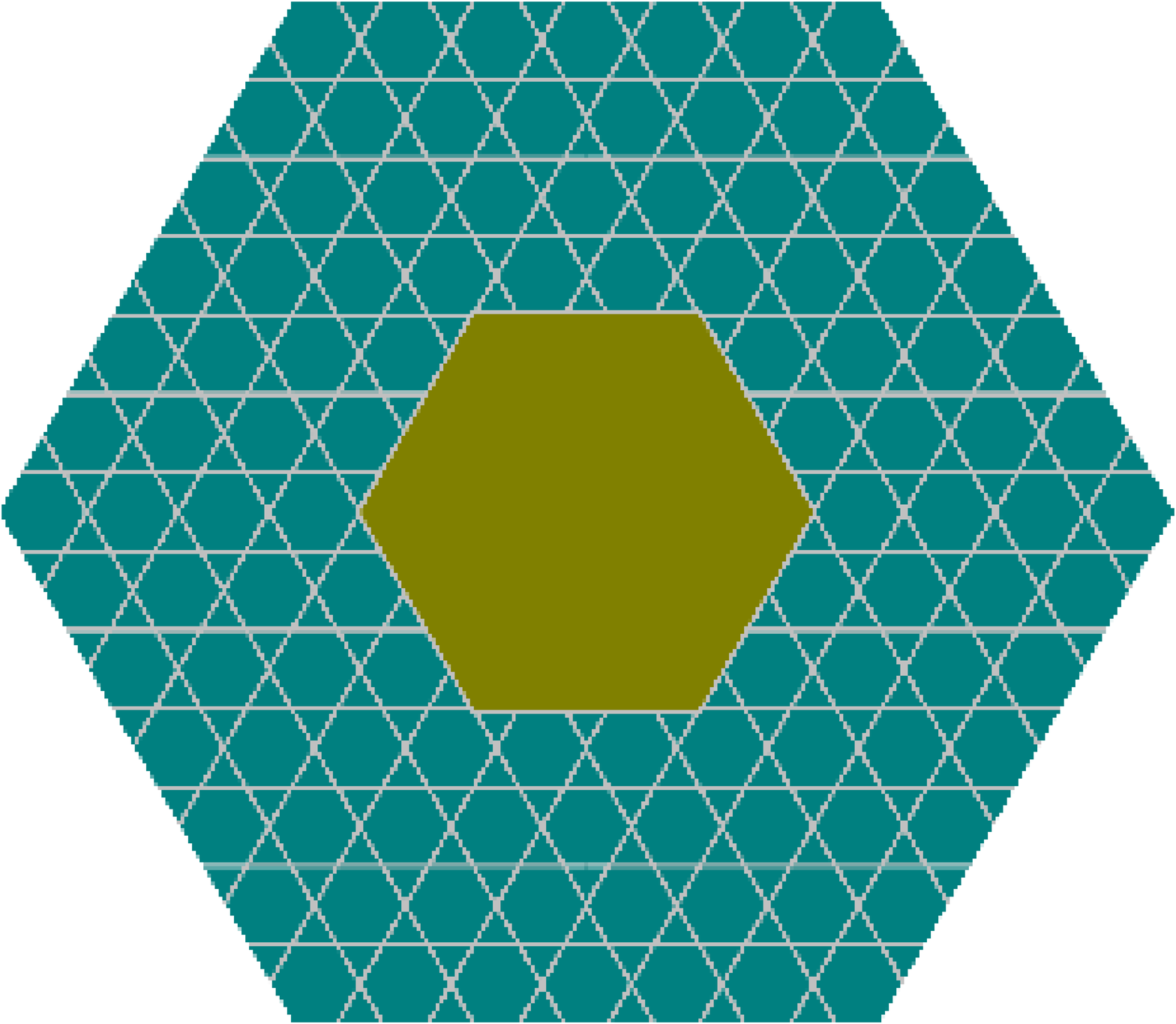}\hfill
\caption{\label{fig:Kagome}(a) Fundamental and (b) fourth core mode of 19-cell kagome-structured HCPCF (c). Parameters  according to Table \ref{table:kagomeLayouts}.}
\end{figure}

In \cite{COU06} attenuation spectra of large-pitch kagome-structured fibers have been measured experimentally. Fig. \ref{fig:Kagome}(a), (b) show the first and fourth fundamental core mode of such a 19-cell kagome-structured fiber. The corresponding layout is given in Fig.\ref{fig:Kagome}(c). 
In this section we compute attenuation spectra numerically. Since we only take into account radiation losses we do not expect quantitative agreement with experimental measurements. Discussions about other loss mechanism can be found in \cite{HOL06,COU06}. We fix the geometrical parameters of the fiber and search for leaky eigenmodes in a wavelength interval. In the experiment core modes could be excited selectively. Here we also use the imaginary part of the fundamental core mode for the attenuation spectra. A small imaginary part then corresponds to low losses and therefore high transmission. Furthermore we look at the confinement of the computed modes. Therefore we compute the energy flux of the mode within $E_{\mathrm{core}}$ and outside $E_{\mathrm{strut}}$ the hollow core. A well confined mode then has a confinement $\frac{E_{{\mathrm{core}}}}{E_{\mathrm{strut}}}$ close to $1$. We expect that well confined modes have small losses.
For simulation we use layouts corresponding to \cite{COU06} shown in Fig. \ref{fig:hcpcf}(b) and \ref{fig:Kagome}(c) with parameters given in Table \ref{table:kagomeLayouts}.
\begin{table}
\centering
\begin{tabular}{lcccc}
Layout & Hollow-core & Pitch $\Lambda$ [$\mu$m]& Strut Width [$\mu$m] \\
\hline
A & 19-cell & 10.9 & 0.51 \\
B & 1-cell & 11.8 & 0.67 \\
\end{tabular}
\caption{\label{table:kagomeLayouts}Layouts of kagome-structured fibers used for mode computation.}
\end{table}

\begin{figure}[ht]
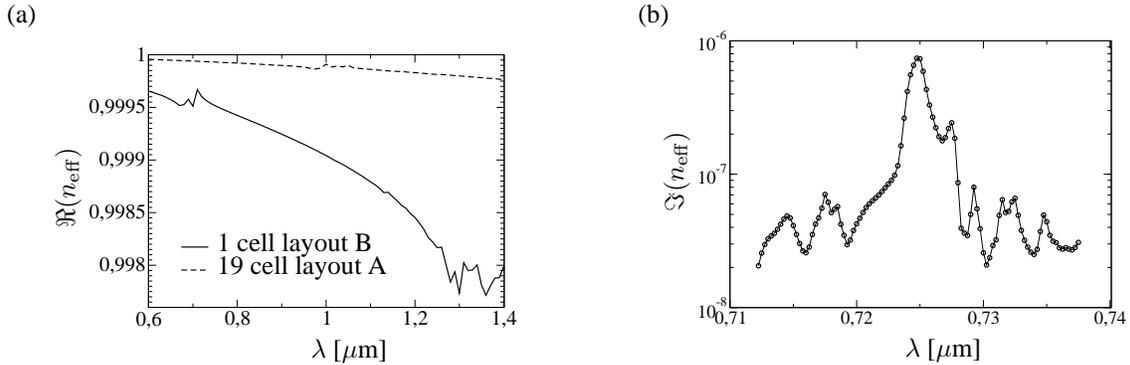

\psfrag{imag}{$\Im({n_{\mathrm{eff}}})$}
\psfrag{real}{$\Re({n_{\mathrm{eff}}})$}
\psfrag{lambda}{$\lambda$ [$\mu$m]}
\psfrag{1cell}{1 cell layout B}
\psfrag{19cell}{19 cell layout A}
(a)\hspace{8cm}(b)\vspace{0mm}\\
\phantom{........}
\includegraphics[width=6cm]{fig/pitch10_9real.eps}\hfill
\includegraphics[width=6cm]{fig/resonance.eps}\hfill
\caption{\label{fig:kagomeReal}(a) Real part of effective refractive index of fundamental leaky mode in dependence on wavelength $\lambda$ for 19-cell and 1-cell kagome-structured fiber. Parameters: see Table \ref{table:kagomeLayouts} - (b) zoom into Fig. \ref{fig:kagomeImag}(a).}
\end{figure}

\begin{figure}[ht]
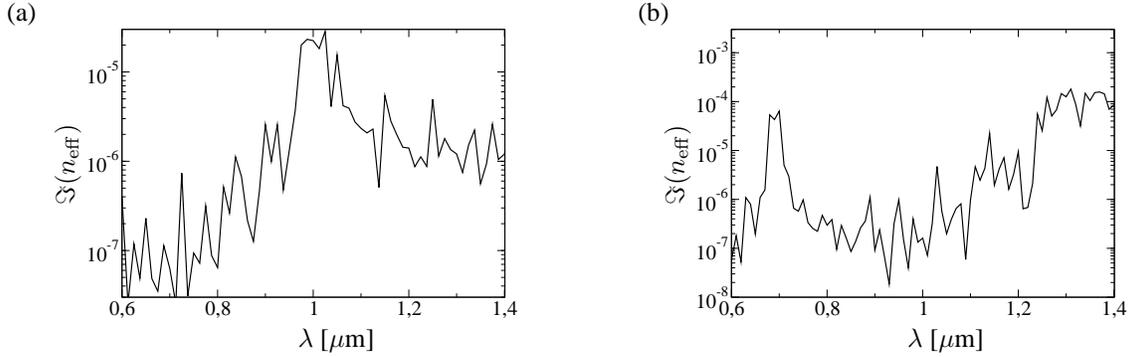

\psfrag{imag}{$\Im({n_{\mathrm{eff}}})$}
\psfrag{lambda}{$\lambda$ [$\mu$m]}
(a)\hspace{8cm}(b)\vspace{0mm}\\
\phantom{........}
\includegraphics[width=6cm]{fig/pitch10_9imag.eps}\hfill
\includegraphics[width=6cm]{fig/cell1.eps}\hfill
\caption{\label{fig:kagomeImag}Imaginary part of effective refractive index of fundamental leaky mode in dependence on wavelength $\lambda$ for (a) 19-cell and (b) 1-cell kagome-structured fiber. Parameters: see Table \ref{table:kagomeLayouts}}
\end{figure}

\begin{figure}
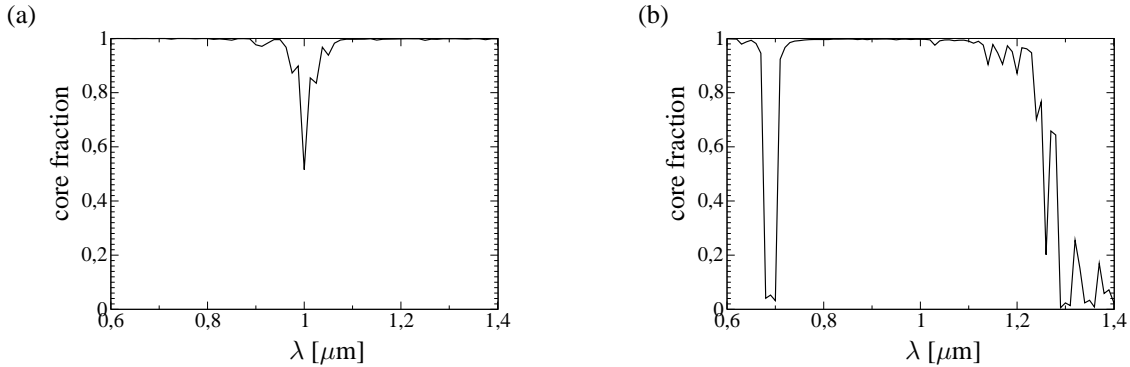

\psfrag{lambda}{$\lambda$ [$\mu$m]}
\psfrag{frac}{core fraction}
(a)\hspace{8cm}(b)\vspace{0mm}\\
\phantom{........}\includegraphics[width=6cm]{fig/cell19core.eps}\hfill
\includegraphics[width=6cm]{fig/cell1core.eps}\hfill\\
\caption{\label{fig:KagomeBand}(a) Fraction of electric field intensity of fundamental leaky mode located inside hollow core in dependence on wavelength $\lambda$ for (a) 19-cell and (b) 1-cell kagome-structured fiber. Parameters: see Table \ref{table:kagomeLayouts}}
\end{figure}

Fig. \ref{fig:kagomeReal}(a) shows the real part of the effective refractive index of the fundamental mode in dependence on the wavelength. Since the core is filled with air it is close to $1$ and hardly changes. Fig. \ref{fig:kagomeImag} shows the imaginary part of the eigenvalues which is proportional to the losses according to Eq. (\ref{eq:dampk_z}). For the 19-cell fiber we find a region of low transmission in the wavelength interval 950nm--1050nm. In \cite{COU06} a region of low transmission spans from 850nm--1050nm.
For the 1-cell fiber we find a peak of high attenuation at $\lambda=700\,$nm and high attenuation from 1200nm-1400nm. A dip in the transmission spectrum can also be found in the corresponding experimental measurement at 630nm. The range of the low transmission band however differs from the simulated. Experimentally it reaches from 800nm-1250nm. An important loss mechanism is coupling of the fundamental mode to interface modes at the glass air interfaces of the fiber \cite{COU06,HOL06}. This is not taken into account in our simulations and could explain the mentioned disagreement. The regions of very high attenuation correspond to poorly confinement modes shown in Fig. \ref{fig:KagomeBand}. 

We notice that both attenuation spectra are very noisy. Since our computed results have converged we assume that this is no numerical artifact. To further investigate the large number of local extrema we zoom into the 19-cell spectrum from 0.715nm--0.735nm where a very large local peak can be seen, see Fig. \ref{fig:kagomeReal}(b). To explain the resonance in the attenuation spectrum at 725nm we look at the field distribution within the kagome structure.

\begin{figure}
(a)\hspace{8.5cm}(b)\\
\includegraphics[width=6cm]{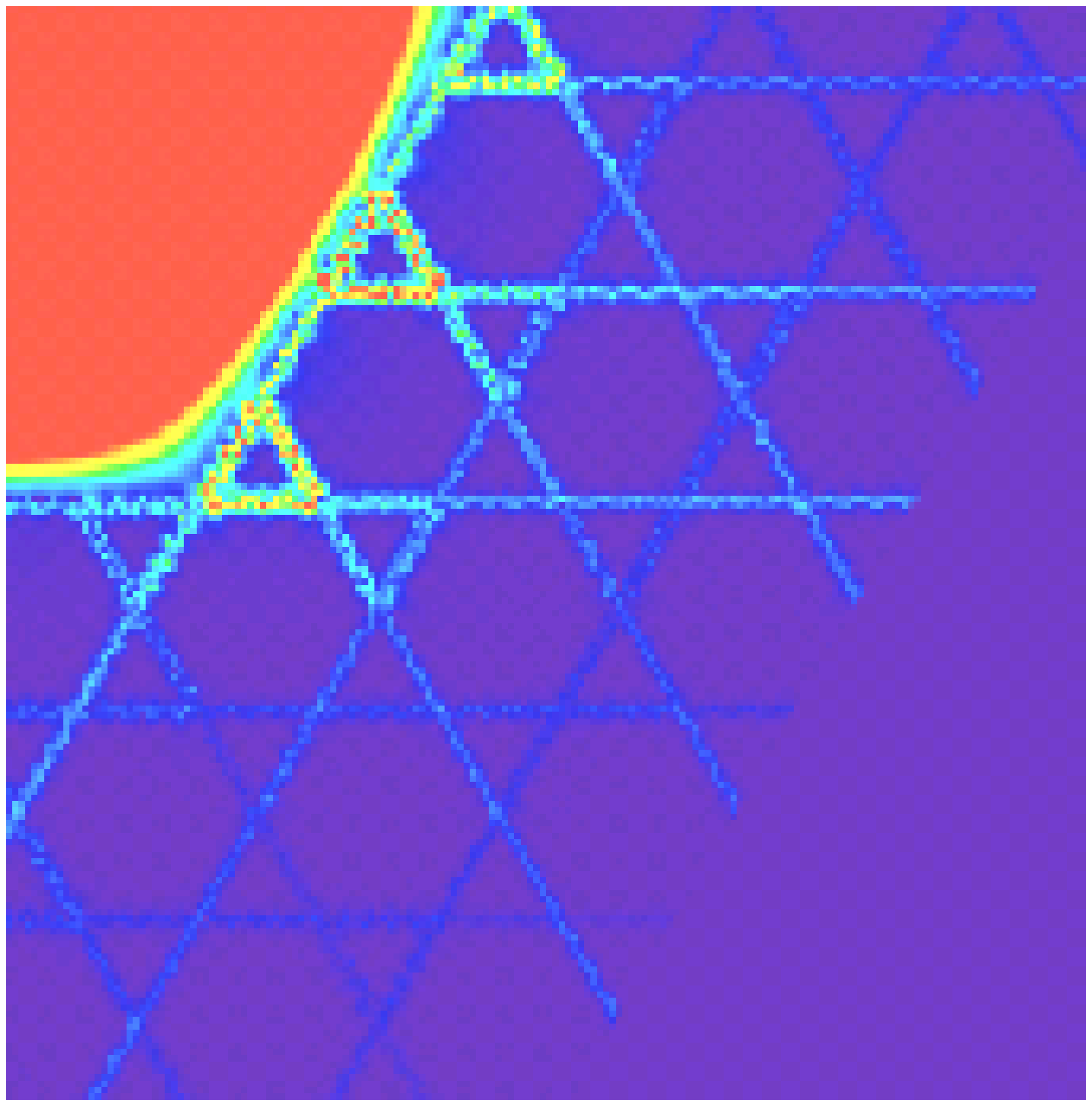}\hfill
\includegraphics[width=6cm]{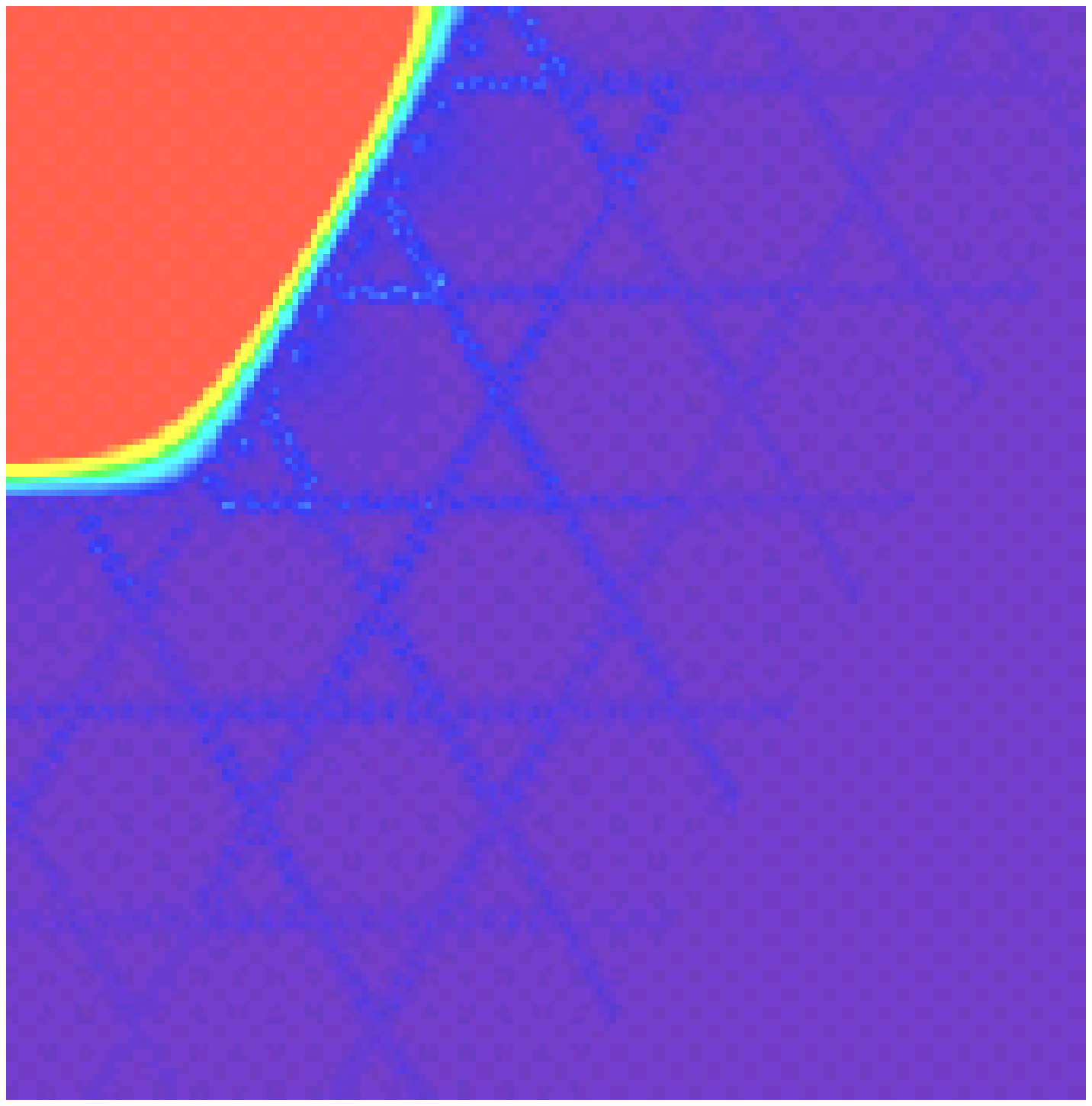}\hfill\\
\caption{\label{fig:kagomeModes}Pseudo color image of intensity of fundamental core mode for (a) $\lambda=725\,$nm and (b) $\lambda=719.25\,$nm. Parameters: Layout A, Table \ref{table:kagomeLayouts}.}
\end{figure}
Fig. \ref{fig:kagomeModes} shows the fundamental eigenmode for $\lambda=719.25\,$nm where we find very low attenuation and for $\lambda=725\,$nm at the local peak attenuation. The intensity of the eigenmodes in Fig. \ref{fig:kagomeModes}(a) and (b) is shown for the same pseudo color range. The field distribution for $\lambda=725\,$nm is more intense in the glass struts which connect the core and the cladding of the fiber. Light from the fundamental core mode is coupled much stronger into the first neighbouring triangles of the kagome structure for $\lambda=725\,$nm. They could be seen as resonators coupled to the hollow core and being excited by the core mode. Better coupling then leads to more light leaving the core and therefore higher attenuation. The coupling into the complicated kagome structure depends very sensitively on the wavelength and the shape of the mode.

Finally Fig. \ref{fig:KagomeBand}(a) shows the confinement of the fundamental mode in dependence on the wavelength.  Regions of low loss correspond to regions of high $\Im(n_{\mathrm{eff}})$, compare Fig. \ref{fig:kagomeImag}.

\section{Conclusion}
We have investigated radiation losses in photonic crystal fibers. We have shown that with our FEM analysis we can efficiently determine the complex eigenvalues to a very high accuracy. The finite size of the photonic crystal fibers was taken into account by transparent boundary conditions realized with the PML method.

Computing leaky propagation modes in HCPCFs eigenvalues with real and imaginary part differing by over 10 orders of magnitude were found. The relative error of the imaginary part was thereby about 7 orders of magnitude larger than the relative error of the real part. In order to precisely compute the much smaller imaginary part of the eigenvalues special techniques were implemented and applied to 2 different types of HCPCFs, namely HCPCFs with hexagonal cladding cells and kagome-structured fibers. A goal-oriented error estimator was introduced which focused on the accurate computation of a target functional, in our case the imaginary part of the eigenvalue. After each computation on a refinement level this error estimator was used to adaptively refine the grid.
In a convergence analysis it was shown that the goal-oriented error estimator led to faster convergence of the quantity of interest compared to uniform refinement or standard field energy based refinement strategies. Also the usage of high order finite elements significantly increased the accuracy of the imaginary part. 

Since the imaginary part of a fiber could be computed in about 10 minutes with a relative error smaller than $10^{-3}$ it was possible to use the finite element method to automatically optimize a fiber design with respect to pitch, strut thickness, cladding meniscus radius and core surround thickness in order to minimize radiation losses. Furthermore attenuation spectra of a 1-cell and a 19-cell kagome-structured fiber were computed and compared to experimental results. It was shown that high losses were connected to poorly confined modes within the hollow-core. Furthermore the appearance of a large number of local minima and maxima in the attenuation spectra was explained by analyzing the intensity distribution of the core modes.

\begin{acknowledgement}
We acknowledge support by the DFG within SPP 1113 under contract number BU 1859/1-1. We thank Ronald Holzl{\"o}hner and John Roberts for fruitful discussions.
\end{acknowledgement}
\bibliography{/home/numerik/bzfpompl/myBib}

\begin{thebibliography}{10}

\bibitem{JOA95}
J.~D. Joannopoulos, R.~D. Meade, and J.~N. Winn.
\newblock {\em {P}hotonic crystals: molding the flow of light}.
\newblock Princeton University Press, 1. edition, 1995.

\bibitem{CRE99}
R.F. Cregan, B.~J. Mangan, J.C. Knight, P.~St.~J. Russel, P.~J. Roberts, and
  D.C. Allan.
\newblock Single-mode photonic band gap guidance of light in air.
\newblock {\em Science}, 285(5433):1537--1539, 1999.

\bibitem{RUS03}
P.~St.~J. Russell.
\newblock Photonic crystal fibers.
\newblock {\em Science}, 299(5605):358--362, 2003.

\bibitem{BEN02}
F.~Benabid, J.~C. Knight, G.~Antonopoulos, and P.~St.~J. Russel.
\newblock Stimulated raman scattering in hydrogen-filled hollow-core photonic
  crystal fibers.
\newblock {\em Science}, 298(5592):399--402, 2002.

\bibitem{COU06}
F.~Couny, F.~Benabid, and P.S. Light.
\newblock Large-pitch kagome structured hollow-core photonic crystal fiber.
\newblock {\em Optics Letters}, 31(24):3574--3576, 2006.

\bibitem{POM07}
J.~Pomplun, R.~Holzl{\"o}hner, S.~Burger, L.~Zschiedrich, and F.~Schmidt.
\newblock {FEM} investigation of leaky modes in hollow core photonic crystal
  fibers.
\newblock volume 6480, page 64800M. Proc. SPIE, 2007.

\bibitem{Bienstman:06a}
P.~Bienstman, S.~Selleri, and L.~Rosa.
\newblock {M}odelling lossy photonics wires: a mode solver comparison.
\newblock page~5. Proc. OWTNM 05, 2006.

\bibitem{Burger2005a}
S.~Burger, R.~Klose, A.~Sch\"adle, and F.~Schmidt and L.~Zschiedrich.
\newblock {FEM} modelling of 3{D} photonic crystals and photonic crystal
  waveguides.
\newblock In Y.~Sidorin and C.~A. W\"achter, editors, {\em Integrated Optics:
  Devices, Materials, and Technologies IX}, volume 5728, pages 164--173. Proc.
  SPIE, 2005.

\bibitem{Burger2006b}
S.~Burger, R.~Klose, A.~Sch\"adle, F.~Schmidt, and L.~Zschiedrich.
\newblock Adaptive {FEM} solver for the computation of electromagnetic
  eigenmodes in 3d photonic crystal structures.
\newblock In A.~M. Anile, G.~Ali, and G.~Mascali, editors, {\em Scientific
  Computing in Electrical Engineering}, pages 169--175. Springer Verlag, 2006.

\bibitem{Burger2005w}
S.~Burger, L.~Zschiedrich, R.~Klose, A.~Sch\"adle, F.~Schmidt, C.~Enkrich,
  S.~Linden, M.~Wegener, and C.~M. Soukoulis.
\newblock Numerical investigation of light scattering off split-ring
  resonators.
\newblock In T.~Szoplik, E.~{\"O}zbay, C.~M. Soukoulis, and N.~I. Zheludev,
  editors, {\em Metamaterials}, volume 5955, pages 18--26. Proc. SPIE, 2005.

\bibitem{Enkrich2005a}
C.~Enkrich, M.~Wegener, S.~Linden, S.~Burger, L.~Zschiedrich, F.~Schmidt,
  C.~Zhou, T.~Koschny, and C.~M. Soukoulis.
\newblock Magnetic metamaterials at telecommunication and visible frequencies.
\newblock {\em Phys. Rev. Lett.}, 95:203901, 2005.

\bibitem{HOL06}
R.~Holzl{\"o}hner, S.~Burger, P.~J. Roberts, and J.~Pomplun.
\newblock Efficient optimization of hollow-core photonic crystal fiber design
  using the finite-element method.
\newblock {\em Journal of the European Optical Society}, 1(06011), 2006.

\bibitem{Kalkbrenner2005a}
T.~Kalkbrenner, U.~H{\aa}kanson, A.~Sch\"adle, S.~Burger, C.~Henkel, and
  V.~Sandoghdar.
\newblock Optical microscopy using the spectral modifications of a
  nano-antenna.
\newblock {\em Phys. Rev. Lett.}, 95:200801, 2005.

\bibitem{LindenEDKZKSBSW06}
S.~Linden, C.~Enkrich, G.~Dolling, M.~W. Klein, J.~Zhou, T.~Koschny, C.~M.
  Soukoulis, S.~Burger, F.~Schmidt, , and M.~Wegener.
\newblock Photonic metamaterials: Magnetism at optical frequencies.
\newblock {\em IEEE Journal of Selected Topics in Quantum Electronics},
  12:1097--1105, 2006.

\bibitem{ZSC07}
L.~Zschiedrich, S.~Burger, J.~Pomplun, and F.~Schmidt.
\newblock {G}oal {O}riented {A}daptive {F}inite {E}lement {M}ethod for the
  {P}recise {S}imulation of {O}ptical {C}omponents.
\newblock volume 6475, page 64750H. Proc. SPIE, 2007.

\bibitem{MON03}
Peter Monk.
\newblock {\em Finite Element Methods for Maxwell's Equations}.
\newblock Oxford University Press, 2003.

\bibitem{BerPML}
J.~B\'erenger.
\newblock A perfectly matched layer for the absorption of electromagnetic
  waves.
\newblock {\em J. Comput. Phys.}, 114(2):185--200, 1994.

\bibitem{Zschiedrich2005a}
L.~Zschiedrich, S.~Burger, R.~Klose, A.~Sch\"adle, and F.~Schmidt.
\newblock Jcmmode: an adaptive finite element solver for the computation of
  leaky modes.
\newblock In Y.~Sidorin and C.~A. W\"achter, editors, {\em Integrated Optics:
  Devices, Materials, and Technologies IX}, volume 5728, pages 192--202. Proc.
  SPIE, 2005.

\bibitem{CUC02}
A.~Cucinotta, S.~Selleri, and L.~Vincetti.
\newblock {H}oley fiber analysis through the finite-element method.
\newblock {\em IEEE Phot. Technol. Lett.}, 14:1530--1532, 2002.

\bibitem{BRE00}
F.~Brechet, J.~Marcou, D.~Pagnoux, and P.~Roy.
\newblock Complete analysis of the characteristics of propagation into photonic
  crystal fibers by the finite-element method.
\newblock {\em Opt. Fiber. Technol.}, 6:181--191, 2000.

\bibitem{Becker:01a}
R.~Becker and R.~Rannacher.
\newblock An optimal control approach to a posteriori error estimation in
  finite element methods.
\newblock In A.~Iserles, editor, {\em Acta Numerica 2000}, pages 1--102.
  Cambridge University Press.

\bibitem{Maerz2004a}
R.~M\"arz, S.~Burger, S.~Golka, A.~Forchel, C.~Herrmann, C.~Jamois,
  D.~Michaelis, and K.~Wandel.
\newblock Planar high index-contrast photonic crystals for telecom
  applications.
\newblock In K.~Busch et~al., editor, {\em Photonic Crystals - Advances in
  Design, Fabrication and Characterization}, pages 308--329. Wiley-VCH, 2004.

\end{thebibliography}
\bibliographystyle{unsrt}

\end{document}